\documentclass[aip, pre, amsmath, amssymb, reprint,onecolumn,superscriptaddress,showpacs]{revtex4-1}
\usepackage{graphicx}
\usepackage{amsmath,amsthm,amsfonts,amscd}

\renewcommand{\vec}[1]{\mbox{\boldmath$#1$}}

\usepackage{color}
\usepackage{rotating}
\usepackage{placeins}
\definecolor{orange}{rgb}{.9,.3,0}

\usepackage{float}
\usepackage{calc}

\begin{document}

\title{Extreme-value statistics from Lagrangian convex hull analysis for homogeneous turbulent Boussinesq convection and MHD convection}
                                            
\author{J. Pratt}
\email[]{j.l.pratt@exeter.ac.uk}
\affiliation{Astrophysics Group, University of Exeter, Exeter, EX4 4QL, United Kingdom}
\author{A. Busse}
\affiliation{School of Engineering, University of Glasgow, Glasgow G12 8QQ, United Kingdom}
\author{W.-C. M\"uller}
\affiliation{Center for Astronomy and Astrophysics, ER 3-2, TU Berlin, Hardenbergstr. 36, 10623 Berlin, Germany}
\author{N.W. Watkins}
\affiliation{Centre for Fusion, Space and Astrophysics, Physics Department, University of Warwick, Coventry, CV4 7AL, United Kingdom}
\affiliation{Institut f\"ur Physik und Astronomie, Universit\"at Potsdam, Campus Golm, Haus 28, Karl-Liebknecht-Strasse
24/25, 14476 Potsdam-Golm, Germany}
\affiliation{Centre for the Analysis of Time Series, London School of Economics, London, United Kingdom}
\affiliation{Faculty of Mathematics, Computing and Technology, Open University, Milton Keynes, United Kingdom}
\author{S.C. Chapman}
\affiliation{Centre for Fusion, Space and Astrophysics, Physics Department, University of Warwick, Coventry, CV4 7AL, United Kingdom}
\affiliation{Max-Planck-Institut  f\"ur Physik komplexer Systeme, 01187 Dresden, Germany}
\affiliation{Department of Mathematics and Statistics, University of Tromso, Norway}
\date{\today}

\begin{abstract}
We investigate the utility of the convex hull of many Lagrangian tracers
to analyze transport properties of turbulent flows with different anisotropy.
In direct numerical simulations of statistically homogeneous and stationary Navier-Stokes
turbulence, neutral fluid Boussinesq convection, and MHD Boussinesq
convection a comparison with Lagrangian pair dispersion shows that convex hull statistics 
capture the asymptotic dispersive behavior of a large group of passive tracer particles.
Moreover, convex hull analysis
provides additional information on the sub-ensemble of tracers that on average disperse most efficiently 
in the form of extreme value statistics and flow anisotropy via the geometric properties of the convex hulls.  
We use the convex hull surface geometry
to examine the anisotropy that occurs in turbulent
convection.  Applying extreme value theory, we show that the maximal
square extensions of convex hull vertices are well described by a
classic extreme value distribution, the Gumbel distribution.  During
turbulent convection, intermittent convective plumes
grow and accelerate the dispersion of Lagrangian tracers. 
Convex hull analysis yields information
that supplements standard Lagrangian analysis of coherent turbulent structures and their influence on 
the global statistics of the flow.  
\end{abstract}


\maketitle

\vspace{5mm}

\section{Introduction}

Turbulent transport governs the spreading of contaminants in the
environment, mixing of chemical constituents in combustion engines or
in stellar interiors, accretion in proto-stellar molecular clouds,
acceleration of cosmic rays, and escape of hot particles from fusion
machines.  Because of its wide relevance, a fundamental characterization
of the dispersive properties of turbulent flows is of practical interest to
physicists and engineers.  Here we examine the broadly relevant case
of dispersion of Lagrangian tracer particles in statistically homogeneous but not necessarily isotropic 
turbulence.  

The Lagrangian viewpoint is particularly suited to the investigation
of transport in turbulent fluids.  A Lagrangian description of turbulence 
is based on following the paths of passive tracer particles in a turbulent 
flow. 
Single-particle diffusion, as originally addressed by
Taylor \cite{taylor:turbdiff}, provides a basic characterization of a flow's transport properties\cite{falkovich_gawedzki_vergassola:review}.
A more complete characterization of the turbulent transport has conventionally been formed from the relative dispersion of two, three, or
four particles \citep{hackl2011multi,sawford2013gaussian,luthi2007lagrangian,toschibodenrev,xu2008evolution,lacasce2008statistics,yeungreview,pumir2000geometry,lin2013lagrangian}. 
However, in astrophysical environments where the effects of magnetic fields, 
rotation, or gravity are often significant, the more complex nature of statistically 
anisotropic or even inhomogenous nonlinear dynamics warrants additional examination.
Dispersion in dynamically anisotropic systems such as vigorously
convecting flows
\citep{shearbursts,mazzitelli2014pair,maeder2008convective,brun2011modeling,leprovost2006self}
where preferred directions exist and spatially coherent, persistent
structures like convective plumes can form, motivates the present consideration of a
complementary diagnostic based on a different Lagrangian concept: the
convex hull\citep{efron65} of a $n$-particle group ($n\gg 4)$.

The convex hull is the smallest convex polygon that encloses a group
of particles; two dimensional convex hulls are pictured in
FIG.~\ref{hull}.  Convex hull analysis of turbulent dispersion is
similar in spirit to following a drop of dye as it spreads in a fluid,
or following a puff of smoke as it spreads in the air, both classical
fluid dynamics problems
\citep{gifsmokepuffs,richardson1926atmospheric,elder1959dispersion}.
A large group of tracer particles can be marked, similarly to adding a
drop of dye to a fluid flow, so that the same particles can be
identified at all later times.  Using the convex hull, a size for the
group of tracer particles that were marked can be calculated at each
time.

The convex hull yields statistical information about a class of
Lagrangian particles that is not equivalent to pre-selected tracer
particle groups like particle pairs or tetrads. These standard
Lagrangian multi-particle statistics represent a fixed and unique
structural relationship between specific tracer particles. The
evolution of particle-pair structures, expressed e.g. as separation and
orientation, is analyzed as the pair of particles is advected by the
fluid. In contrast, the convex hull does not establish a unique link
between the tracers that generate it, but continuously selects from a
predefined group, based on which tracer particles have
ventured furthest from the geometrical center of the ensemble.  Unlike
particle pairs or tetrads, the particles that constitute the convex
hull are dynamically changing.  The definition of the convex hull thus
corresponds to a filtering based on the entire dynamical past of each
particle in the group. 
The convex hull captures the extremes of the excursions of a
group of particles, information relevant to the non-Gaussian aspects
of the dynamics.  The behavior of particles that do not exhibit the fastest dispersion is
filtered by the convex hull, allowing a classification of particle dynamics with regard
to their dispersion efficiency. 
In this work we begin to explore this link to extreme value theory,
which has the potential to provide new physical insight for turbulent
diffusion.  
The dynamical relation between the Lagrangian particle population forming the convex
hull and the bulk ensemble of tracer particles enclosed by it represents another 
aspect of this diagnostic that could be exploited in investigations of turbulent 
structure formation.

In recent years, convex hull calculations have been used
to study diverse topics such as the size of spreading
GPS-enabled drifters moving on the surfaces of lakes and rivers
\citep{lakes,spencer2014quantifying}, star-forming clusters \citep{schmeja}, forest fires \citep{forestfire}, proteins \citep{li08,millett2013identifying}, or clusters of contaminant particles \citep{dietzel2013numerical}.
Studies of the relationships
between random walks, anomalous diffusion, extreme statistics and convex
hulls have been motivated by animal home ranges \citep{eco2006,ecoPRL,maj2010,lukovic2013area,vander2013trophic,dumonteil2013spatial,collyer2015habitat}.  Convex hulls have also been used to study analytical statistics of Burgers turbulence by analogy with Brownian motion \citep{avellaneda1995statistical,bertoin2001some,chupeau2015convex}.
 
MHD turbulence \citep{busse_hoho,homann2014structures} and turbulence
during hydrodynamic convection
\citep{schumacher09,schu2008,forster2007parameterization} are areas
where statistical analysis of Lagrangian  particles has begun to
be applied only recently.  This work presents new Lagrangian results
from three-dimensional direct numerical simulations of turbulent MHD
Boussinesq convection, and compares them with turbulent hydrodynamic
Boussinesq convection and homogeneous isotropic turbulence.  It is
structured as follows.  In Section \ref{section2} we describe the
fluid simulations.  In Sections \ref{section3} we present standard
Lagrangian pair dispersion and
discuss the results of these widely-used statistical tools for
convective flows.  In Section \ref{section4} we describe the convex
hull analysis that we perform on groups of many Lagrangian tracer
particles.  We perform several basic checks on our convex hull
calculations.  We then compare the dispersion curves obtained from
convex hulls of large groups of particles with the
expected scalings for particle-pair dispersion.
  In Section
\ref{section5} we demonstrate how the convex hull can be used to
examine anisotropy.  In Section \ref{section6} we apply extreme value
theory, and show that the maximal square extensions of convex hull
vertices are well described by a classic extreme value distribution,
the Gumbel distribution.  In Section \ref{section7} we summarize the
results of this validation study, and our extreme value statistics.
We discuss the potential uses and benefits of convex hull analysis.

\section{Simulations \label{section2}}

We investigate three different types of turbulent systems: forced
homogeneous isotropic Navier-Stokes turbulence (simulation
NST)\citep{angeladiss09}, Boussinesq convection in a neutral fluid
(simulation HC), and Boussinesq convection in an electrically
conducting fluid (simulation MC)\citep{shearbursts,moll2011}. 
These simulations are not designed for close comparison, but produced for
a broad exploration of the convex hull analysis. In each
of these direct numerical simulations, the equations are solved using
a pseudospectral method in a cubic simulation volume with a side of
length $2\pi$.  The non-dimensional Boussinesq equations for MHD
convection in Alfv\'enic units are
\begin{eqnarray} \label{realbmhdc}
\frac{\partial \vec{\omega} }{\partial t} &-& \nabla \times (\vec{v} \times \vec{\omega} +  \vec{j} \times \vec{B}) 
=  \hat{\nu} \nabla^2 \vec{\omega} - \nabla \theta \times \vec{g}_0
\\
\frac{\partial \vec{B} }{\partial t} &-& \nabla \times (\vec{v} \times \vec{B}) =  \hat{\eta} \nabla^2 \vec{B}
\\ \label{thermeq}
\frac{\partial \theta }{\partial t} &+& (\vec{v} \cdot \nabla) \theta = \hat{\kappa} \nabla^2 \theta -  (\vec{v} \cdot \nabla) T_0\\
&& \nabla \cdot \vec{v}=  \nabla \cdot \vec{B}=0 ~~.
\end{eqnarray}
These equations include the solenoidal velocity field $\vec{v}$,
vorticity $\vec{\omega}=\nabla\times\vec{v}$, magnetic field
$\vec{B}$, and current $\vec{j}=\nabla\times\vec{B}$.
The quantity $\theta$ denotes the temperature fluctuation about a
linear mean temperature profile $T_0(z)$ where $z$ is the direction of
gravity.  In eq. \eqref{thermeq} this mean temperature gradient
provides the convective drive of the system.  In
eq. \eqref{realbmhdc}, the term including the temperature fluctuation
$\theta$ is the buoyancy force.  The vector $\vec{g}_0$ is a unit
vector in the direction of gravity.  Three dimensionless parameters
appear in the equations: $\hat{\nu}$, $\hat{\eta}$, and
$\hat{\kappa}$.  They derive from the kinematic viscosity $\nu$, the
magnetic diffusivity $\eta$, and thermal diffusivity $\kappa$.

For simulation HC, the magnetic field $\vec{B}$ is set to zero.  For
simulation NST, both magnetic field terms and temperature terms are
zero.  A fixed time step and a
trapezoidal leapfrog method \citep{kurihara1965use} are used for the
time-integration for simulation NST.  The Boussinesq convection
simulations HC and MC are integrated in time using a low-storage
3rd-order Runge Kutta scheme \citep{will80} and an adaptive time step,
which allows for better time resolution of large fluctuations that
occur during convection.
 
\begin{figure}
\resizebox{3.3in}{!}{\includegraphics{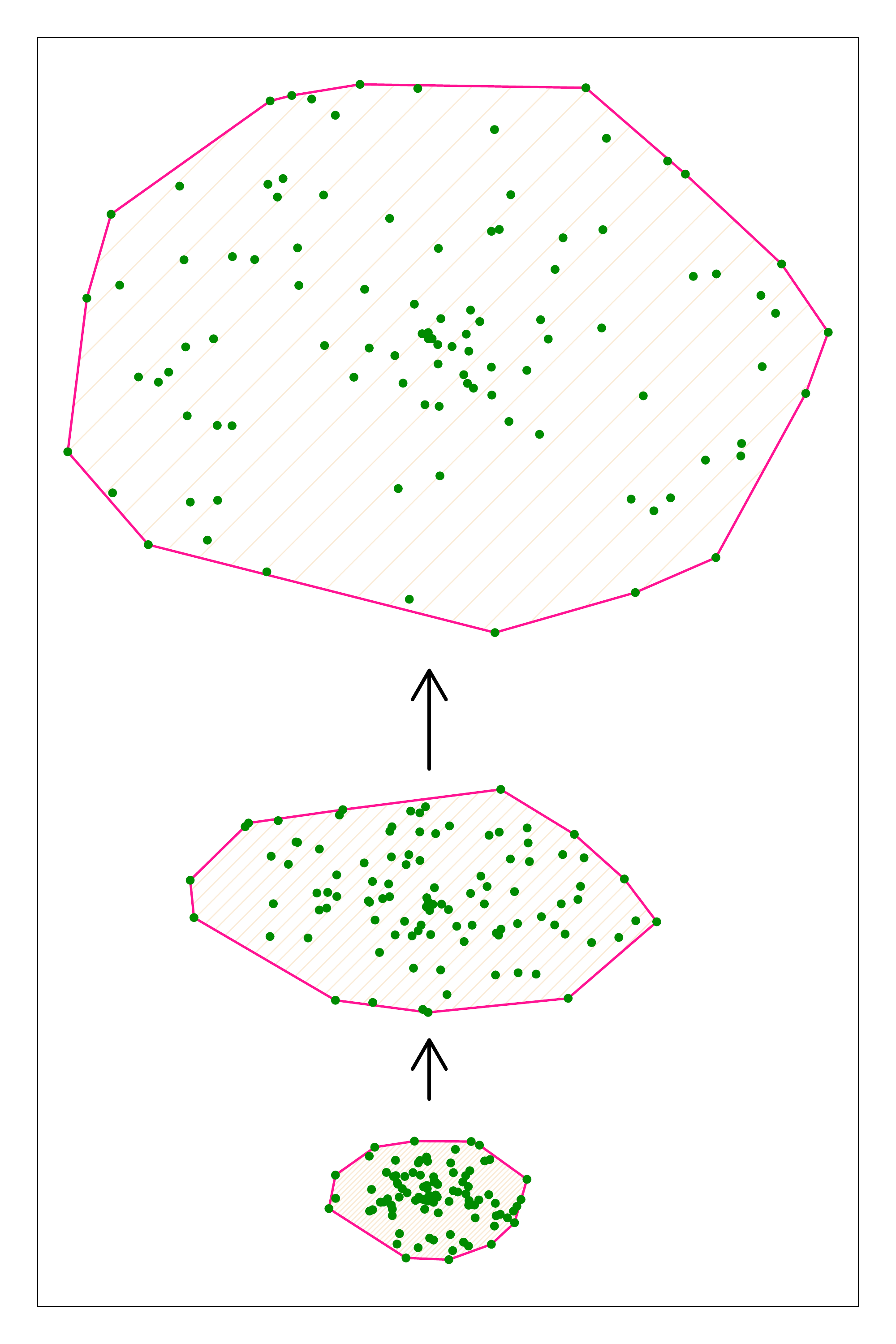}}
\caption{ An illustration of a two dimensional convex hull (solid line) surrounding a group of particles (solid points) as they disperse in time.  The time progression is indicated by arrows, and the particles in each of the three convex hulls shown are the same.  
\label{hull} }
\end{figure}

In this work we discuss turbulent dispersion in an incompressible
fluid, where conservation of volume is a primitive concept.  A volume
of fluid that is convex at an initial time will occupy the same volume
after a period of dynamic development but will generally change its
shape and lose its convexity.  Lagrangian tracer particles that are
contained in the initial volume are marked so that they can be
followed for the entire time of the simulation. At any later time, the
volume of the convex hull of that group of marked particles is
generally not conserved.  This is illustrated in FIG.~\ref{hull} for a
group of particles, and for snapshots taken at three times.  The
growth of surface area and volume are natural concepts for convex
hulls.
 
 A summary of the fundamental parameters that describe each simulation
is given in Table~\ref{simsuma}.  In this table, we define the
Reynolds number to be $\mathsf{Re= \langle E_v^{1/2} L\rangle
/\hat{\nu}} $, where $\mathsf{E_v}= \vec{v}^2/ 2$ is the kinetic
energy, and the brackets indicate a time-average.  We define the
characteristic length scale $\mathsf{L}$ based on the largest-scale
motions of the system in question.  For statistically homogeneous
turbulent convection the characteristic length scale is the
instantaneous temperature gradient length scale $\mathsf{L}=T_*/\nabla
T_0$ where $T_{*}$ is the root-mean-square of temperature fluctuations
and $\nabla T_0$ is the constant vertical mean temperature gradient
\citep{gibert2006high}.  For non-convective statistically homogeneous
turbulent flows, the characteristic length scale is a dimensional
estimate of the size of the largest eddies,
$\mathsf{L}=\mathsf{E_v}^{3/2}/\mathsf{\epsilon_{\mathrm{v}}}$, where
$\mathsf{\epsilon_{\mathrm{v}}}=\hat{\nu} \langle \sum_k k^2 \vec{v}^2
\rangle$ is the time-averaged rate of kinetic energy dissipation.  The
magnetic Reynolds number is defined from the Reynolds number and the
magnetic Prandtl number, i.e. $\mathsf{Re_m}=\mathsf{Pr_m Re}$.  We
measure length in units of the Kolmogorov microscale
$\mathsf{\eta_{kol}=(\hat{\nu}^3/\mathsf{\epsilon_{\mathrm{v}}})^{1/4}}$
and also make use of the Kolmogorov time-scale
$\mathsf{\tau_{\eta}=(\hat{\nu}/\mathsf{\epsilon_{\mathrm{v}}})^{1/2}}$.
The Kolmogorov microscale multiplied by
$\mathsf{k_{\mathrm{max}}}$, the highest wavenumber in the simulation,
is often used to test whether a simulation is adequately resolved on
small spatial scales.  In this work all of the simulations fulfill the
standard criterion based on the Kolmogorov microscale
($\mathsf{k_{\mathrm{max}} \mathsf{\eta_{kol}} >1.5 }$) for adequate
spatial resolution \citep{pope2000turbulent}.  The Reynolds numbers in
Table~\ref{simsuma} are on the order of $10^3$; the Reynolds numbers
and Kolmogorov microscales in Table~\ref{simsuma} are in the same
range as current studies of moderately turbulent flows
\citep[e.g.][]{marino2015helical,bianchi2016evolution}.

\begin{table*}
\caption{Simulation parameters: grid size $N^3$, total number of particles in the simulation $\mathsf{n_p}$ ($10^6$), Reynolds number $\mathsf{Re}$, magnetic Reynolds number $\mathsf{Re_m}$, Prandtl number $\mathsf{Pr}$, magnetic Prandtl number $\mathsf{Pr_m}$,  Rayleigh number $\mathsf{Ra}$, Kolmogorov microscale $\eta_{\mathsf{kol}}$, Kolmogorov time-scale $\tau_{\eta}$, Lagrangian crossing time $\mathsf{LCT}$, average kinetic energy dissipation rate $\epsilon_{\mathrm{v}}$, Alfv\'en ratio $r_{\mathrm{A}}$, average Bolgiano-Obukhov length divided by the height of the simulation volume $\bar{\ell}_{\mathrm{bo}}$, average number of particles per convex hull $\mathsf{n_{pch}}$, number of convex hulls $\mathsf{N_{hulls}}$, initial length scale of convex hull $\ell_{\mathrm{hull}}$.
\label{simsuma}
 }
\begin{tabular}{lcccccccccccccccccccccccccccc}
\hline\hline
                           & $N^3$ & $\mathsf{n_p}$($10^6$)  & $\mathsf{Re}$  & $\mathsf{Re_m}$  
                           & $\mathsf{Pr}$ & $\mathsf{Pr_m}$ & $\mathsf{Ra}$ ($10^5$) &  $\eta_{\mathsf{kol}}$ ($10^{-3}$) & $\tau_{\eta}$($10^{-2}$)
                           & $\mathsf{LCT}$ ($\tau_{\eta})$  &$\epsilon_{\mathrm{v}}$ & $r_{\mathrm{A}}$ & $\bar{\ell}_{\mathrm{bo}}$ &  $\mathsf{n_{pch}}$  
                           &  $\mathsf{N_{hulls}}$  & $\ell_{\mathrm{hull}}(\eta_{\mathsf{kol}})$
\\ \hline
NST                        & $1024^3$ & 3.2           & 2900  & -  & -  & -      & -  & 4.58   & 5.25  & 276 & 0.15 & -       & -  & 24   & 5000  & 27 
 \\ \hline
HC                         & $512^3$ & 1.0           &  1500 & -  &  2.0   &  -   & 5  & 12.6  & 3.97  & 340 & 2.54 & -    & 0.28   &   214    & 2500  & 22
 \\ \hline
MC                         & $512^3$ & 1.0           &  5100  & 7650 &  2.0  & 1.5  &  2.22  & 8.9  &  2.60 & 530 & 4.40 & 1.78  & 0.12 & 48  & 2500 & 30  
\\ \hline\hline
\end{tabular}
\end{table*}

Formulation of boundary conditions for simulations of
turbulent flows is delicate because boundaries strongly influence the
structure and dynamics of the flow.  For homogeneous isotropic
turbulence, it is standard to employ boundary conditions that are
periodic in $x$, $y$, and $z$.  These fully periodic boundary
conditions are used for simulation NST.  For convection simulations the
choice of fully periodic boundary conditions (also called homogeneous
Rayleigh-B\'enard boundary conditions) allows macroscopic elevator
instabilities to form \citep{calzavarini_etal:elevator}.  These
instabilities destroy the natural pattern of the original turbulent
flow field.  The convection simulations discussed in this work use
quasi-periodic rather than fully periodic boundary conditions.  In
quasi-periodic boundary conditions the only additional constraint is
the explicit suppression of mean flows parallel to gravity, which are
removed at each time step.  Because our simulations are
pseudospectral, the mean flow is straightforwardly isolated as the $z$
component of the $\vec{k}=(0,0,0)$ mode in Fourier space, which
corresponds to the volume-averaged velocity in the z-direction.
Quasi-periodic boundary conditions combine the conceptual simplicity
of statistical homogeneity with a physically natural convective
driving of the turbulent flow.  These boundary conditions do not
enforce a large-scale structuring of the turbulent flow, such as the
convection-cell pattern observed when Rayleigh-B\'enard boundary
conditions are used.  In the quasi-periodic simulations presented in
this work, we find no evidence of the macroscopic elevator instability
although we follow the evolution of the flow for long times.
Quasi-periodic boundary conditions allow for direct comparison with
simulations that use fully periodic boundary conditions.

In simulation NST the modes $2.5<k<3.5$ are forced using
 Ornstein-Uhlenbeck processes with a finite time-correlation on the
 order of the autocorrelation time of the velocity field (for further
 details of this forcing method, see \citet{eswaran1988examination}).  The
 convection simulations HC and MC are Boussinesq systems driven solely
 by a constant temperature gradient in the vertical direction.  The
 magnetic field present in simulation MC is generated
 self-consistently by the flow from a small random seed field through
 small-scale dynamo action.  The system is evolved 
 until a statistically stationary state is
 reached.  For Boussinesq convection, a length scale that
 characterizes the scale-dependent importance of convective driving is
 the Bolgiano-Obukhov length,
 {$\ell_{\mathrm{bo}}=\mathsf{\epsilon_{\mathrm{v}}^{5/4}/\epsilon_{\mathrm{T}}^{3/4}}$},
 where $\mathsf{\epsilon_{\mathrm{ T}}}$ is the average rate of
 thermal energy dissipation.  This length scale separates
 convectively-driven scales of the flow {$\ell > \ell_{\mathrm{bo}}$}
 from the range of scales where the temperature fluctuations behave as
 a passive scalar $\ell < \ell_{\mathrm{bo}}$.  In Table~\ref{simsuma}
 this length scale is averaged over the simulation time, normalized to
 the height of the simulation volume, and recorded as
 $\bar{\ell}_{\mathrm{bo}}$.  The table also includes the mean
 Alfv\'en ratio, $r_{\mathrm{A}} = \langle \mathsf{E_v} /\mathsf{E_b}
 \rangle$, the time-average of the kinetic energy divided by the
 magnetic energy $\mathsf{E_b}= \vec{B}^2 / 2$.  In the present
 numerical experiments, Navier-Stokes turbulence displays the weakest
 form of spatial coherence while Boussinesq magneto-convection
 exhibits anisotropy with regard to the direction of
 gravity as well as the occurrence of large-scale spatially-coherent
 structures.  Additionally,  a dynamical anisotropy arises because of
 the presence of magnetic fields.

The positions of Lagrangian tracer particles are initialized in a
homogeneous random distribution at a time when the turbulent flow is
in a statistically stationary steady state. The total number of
particles in the simulation, $\mathsf{n_p}$, is listed in
Table~\ref{simsuma}.  We use at least a million particles for a
$512^3$ grid.  This is a standard spatial density of tracer particles
used to describe homogeneous turbulence
\citep[e.g.][]{biferale2005lagrangian,homann2007lagrangian,busse2007statistics}. The
Lagrangian statistics we produce have been tested and found to be
well-resolved in space and time; we reproduce these statistics with
half the particles.  At each time step the particle velocities are
interpolated from the instantaneous Eulerian velocity field using
either a trilinear (for simulations HC and MC) or tricubic (for
simulation NST) polynomial interpolation scheme.  Particle positions
are calculated by numerical integration of the equations of motion
using a predictor-corrector method.  For the convex hull calculations,
the Lagrangian particle data is resampled at a rate of approximately
$\tau_{\eta}/10$ for simulations NST and HC.  The rate of sampling for
simulation MC was smaller by a factor of $10$, and this was not found to impact the
dispersive results examined here.  Each simulation is run for a sufficient time that Lagrangian
particle pairs have separated, on average, at least by the length of the
simulation volume.  We call this time the Lagrangian crossing time,
$\mathsf{LCT}$, and it is listed in the table in units of the
Kolmogorov time scale.  Lagrangian
particle pair dispersion statistics exhibit a diffusive trend near
this time since the velocity fluctuations over this time and distance
exhibit low correlation. 

\section{Pair Dispersion of Lagrangian tracer particles during homogeneous Boussinesq convection}\label{section3}

\begin{figure}
\resizebox{3.375in}{!}{\includegraphics[angle=90]{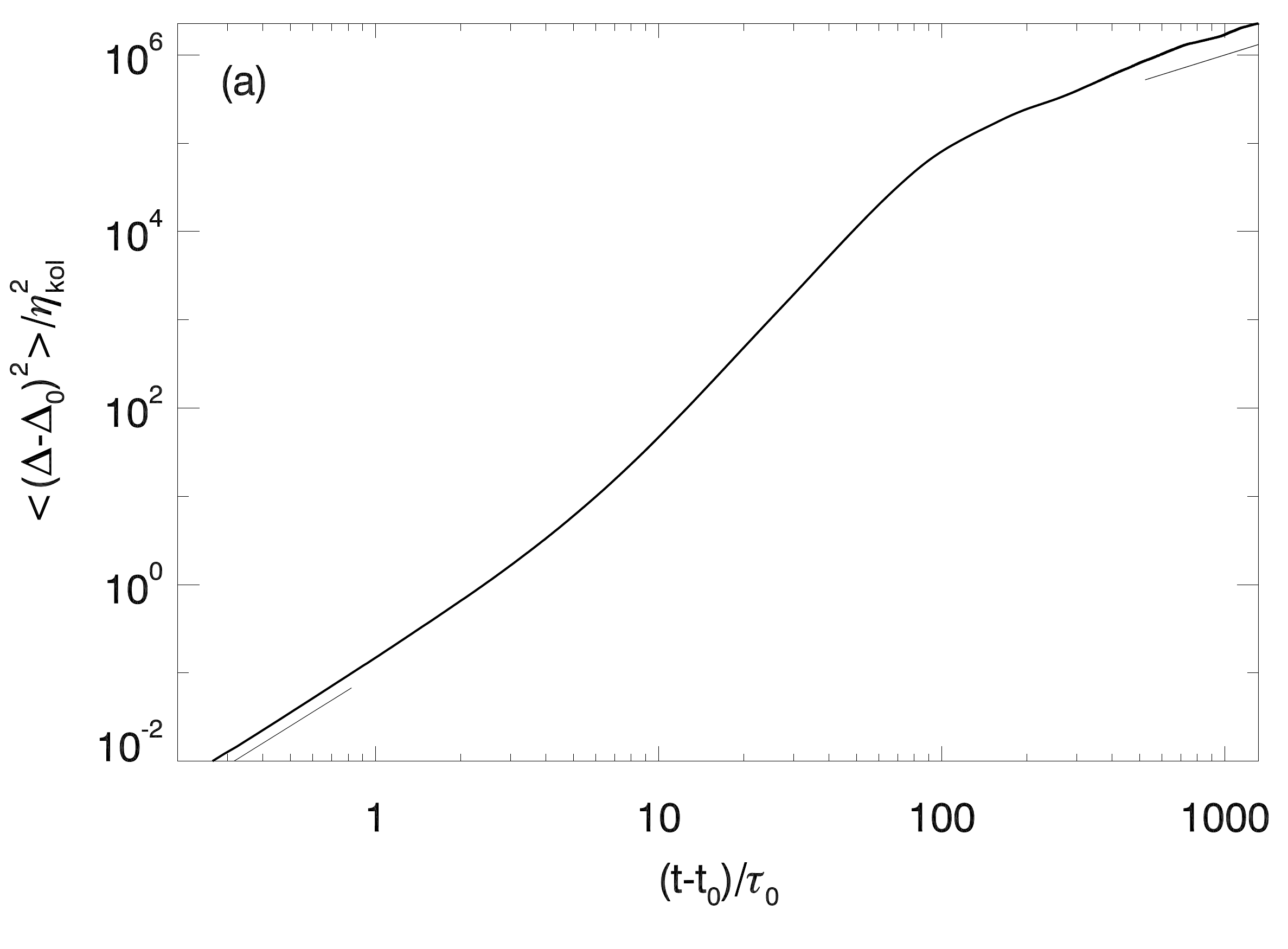}}\resizebox{3.375in}{!}{\includegraphics[angle=90]{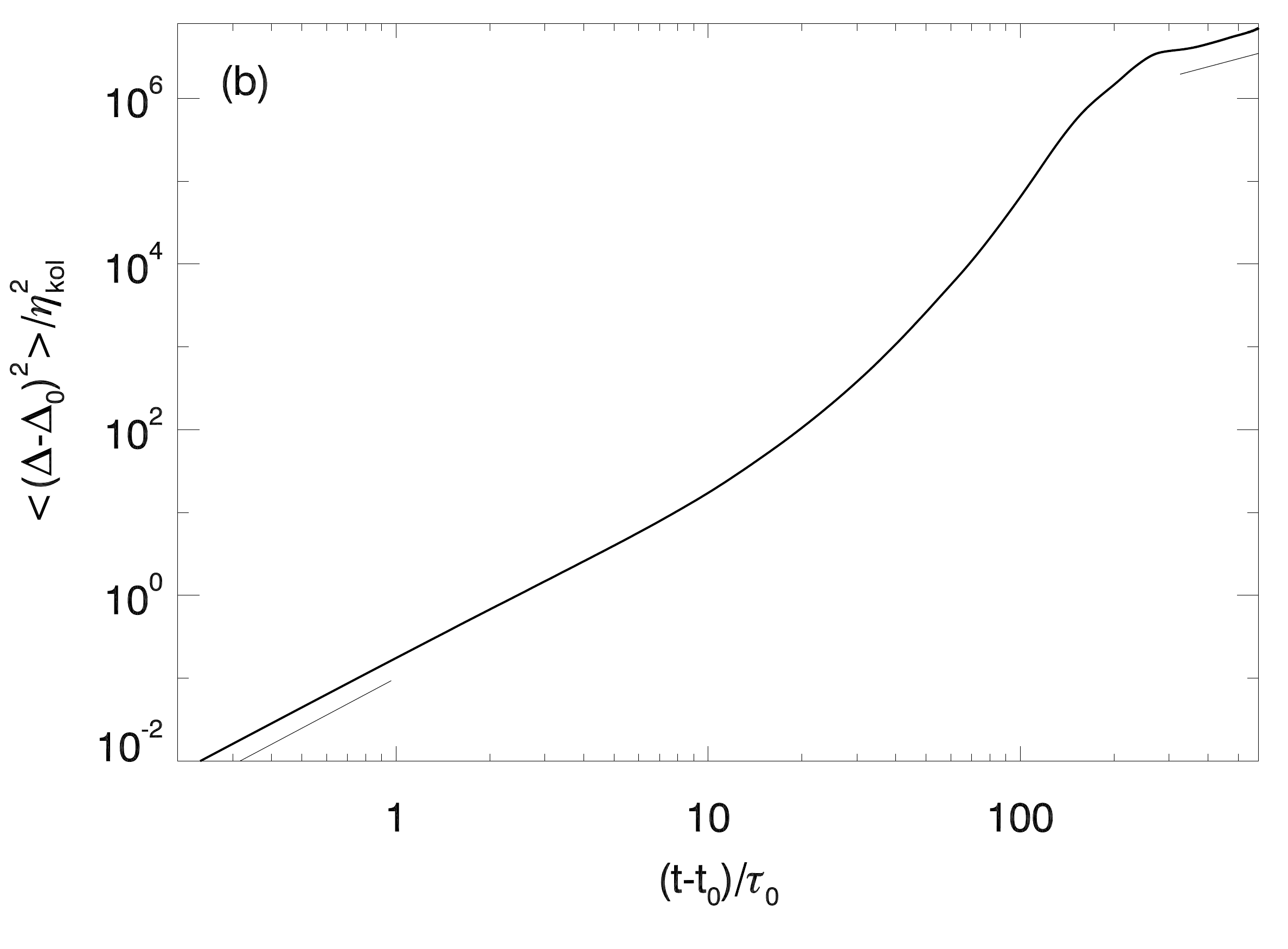}}
\caption{Mean-square of the separation in the direction of gravity for pairs of Lagrangian tracer particles, dispersing in the hydrodynamic convection simulation HC (a) and the MHD convection simulation MC (b).  Particle pairs are initially separated in the direction of gravity by $\Delta_0=\eta_{\mathsf{kol}}$ (HC) and $\Delta_0=1.4\eta_{\mathsf{kol}}$ (MC). Thin solid lines indicate: Batchelor scaling $\sim t^2$ (short timescales) and diffusive scaling $\sim t$ (long timescales). Time and length are given in units of the initial turnover time $\tau_{0}$ and the Kolmogorov microscale $\eta_{\mathsf{kol}}$, respectively.
\label{pairdispfig} }
\end{figure}

This section presents results for particle-pair dispersion during homogeneous
Boussinesq convection for comparison with many-particle dispersion calculated from a convex
hull analysis. For an introduction to the
rich field of Lagrangian particle-pair dispersion, we refer the reader
to the review of \citet{salazar2009two}. In addition to this review, several more recent works\citep{bourgoin:disppheno, thalabard_krstulovic_bec:ctrwdisppheno, bitane_homann_bec:disptimescale} propose new dispersion phenomenologies based on locally ballistic dynamics, an alternative to 
the classical idea of turbulent diffusion exhibiting scale-dependent diffusivity \citep{richardson1926atmospheric}.  Here we briefly recall the
basic argument for scaling regimes of pair dispersion.  For
times short compared with the autocorrelation time of the Lagrangian
velocities, the relative velocity of the particles is approximately
constant.  The mean-squared separation of a pair of Lagrangian
particles is therefore expected to grow quadratically with time for a
short time. This is called the \emph{ballistic} or
\emph{Batchelor} regime. The extent of the ballistic regime is known to depend on the \emph{initial} separation of the particle pair, $\Delta_0$,
due to a finite correlation of $\Delta_0$ and the root-mean-square (RMS) velocity fluctuations on this scale $v_{\Delta_0}$.
Recent theoretical\cite{bourgoin:disppheno,thalabard_krstulovic_bec:ctrwdisppheno,bitane_homann_bec:disptimescale} and experimental\cite{ouellette_etal:pairdispexpmodels,ouellette_etal:pairdispexp}
works make use of a key time scale linked to $\Delta_0$, the initial nonlinear turnover time $\tau_0 \equiv \Delta_0/v_{\Delta_0}$.  In the inertial range of Navier-Stokes turbulence this 
initial turnover time can be estimated as $\tau_0 \sim v_{\Delta_0}^2/(2\epsilon_{\mathrm{v}})$.
For times much larger than the
autocorrelation time of the Lagrangian velocity, the velocities of a
pair of Lagrangian particles are statistically independent.  The
mean-squared separation of a pair of Lagrangian particles is expected
to grow linearly with time.  This is typically
called the \emph{diffusive} regime.  In between the ballistic regime
and the diffusive regime is a period of time where the
mean-squared separation of particle pairs can grow cubically with
time. 
This is typically called the \emph{Richardson-Obukhov regime}.
The temporal separation of the ballistic and Richardson-Obukhov regimes \cite{bitane_homann_bec:disptimescale}
can be estimated by $\tau_0$.
Achieving a clear Richardson-Obukhov regime in direct numerical
simulations depends on the initial separation of particles as well as
the size of the inertial range, and is the subject of current ongoing
research for Navier-Stokes turbulence.  For this reason, and due to the limited extent
of the inertial scaling range that is expected for the Reynolds numbers we obtain,
we make no claims of 
observing a Richardson-Obukhov regime in the present convection simulations.
We compute the initial turnover time 
via a one-dimensional Eulerian kinetic energy spectrum as $\tau_0=(k_0^3 \mathsf{E_v}(k_0))^{-1/2}$ with $k_0=2\pi/\Delta_0$.
Although the moderate Reynolds numbers of the present simulations are far away from values 
where a true inertial range, devoid of influences from largest or smallest scales of the flows
could be realized, for descriptive convenience we will apply this term to the interval of time-scales
between the ballistic and the diffusive regime.
FIG.~\ref{pairdispfig} illustrates the Lagrangian particle-pair
dispersion for simulations HC and MC, both driven with homogeneous
Boussinesq convection characterized by a large Bolgiano-Obukhov
length.  In this figure, thin solid lines indicate 
Batchelor scaling $\sim t^2$ and diffusive scaling $\sim t$, around the shortest and longest timescales,
respectively. For both cases, HC and MC, we have $\Delta_0\simeq \eta$, and thus $\tau_0\simeq \tau_\eta$. Indeed,
both curves deviate from $t^2$-scaling after $\tau_0$. For intermediate times $10 \tau_0 \lesssim (t-t_0) \lesssim 100 \tau_0$, they display a phase of fast separation which eventually levels off toward diffusive dispersion.
The onset of fast pair separation in convection at approximately $10\tau_0$ is delayed compared to 
Navier-Stokes turbulence where it has been observed\cite{bitane_homann_bec:disptimescale} to begin at $(t-t_0)\simeq \tau_0$. 
In a simulation of convection an anisotropy exists
between the direction of the mean temperature gradient and the
direction perpendicular.  The separation of particle pairs evolves
differently in these two directions; the separation of particle pairs
can also evolve differently depending on whether the pair
of particles are initially separated in the direction of the mean
temperature gradient or perpendicular to it.

During Boussinesq convection with large Bolgiano-Obukhov length the
Batchelor regime for pair separations looks similar to randomly forced
hydrodynamic turbulence driven at the large scales, as shown by
e.g. \citet{sawfordreview,yeung2004relative}. During the diffusive
regime, large-scale flow structures associated with large
Bolgiano-Obukhov length Boussinesq convection clearly affect the pair
dispersion curve. The dispersion curve does not look as smooth as the
result obtained from randomly forced hydrodynamic turbulence driven at
the large scales.  This is
not surprising because the separation of the particle pairs has reached sizes
comparable to the large-scale convective plumes. We note that although our convection simulations use
quasi-periodic boundary conditions, FIG.~\ref{pairdispfig} is not
qualitatively different from figure 2 of \citet{schu2008}, which
presents Lagrangian dispersion during Rayleigh-B\'enard convection.
For pair dispersion in simulations HC and MC, extensive averaging over
different flow realizations would be necessary to achieve a perfectly
smooth and universal result, free from the influence of intermittent
plumes or large-scale magnetic structures.

\section{Convex Hull Analysis of Dispersing Tracer Particles \label{section4}}

\subsection{Description of the convex hull calculations}

We seed a number of tracer particles in the simulation volume, which
produces a fixed density of tracer particles.  In simulations NST, HC, and MC the
number of tracer particles and their density is based on the number
needed to produce well-resolved Lagrangian pair dispersion statistics.
A convex hull analysis could potentially make use of a significantly
higher density of tracer particles.  
To calculate a convex hull, we select and mark a group of Lagrangian
tracer particles initially contained in a small cubic sub-volume of
our simulation. 
The 
initial length scale of the group of particles $\ell_{\mathrm{hull}}$
is calculated as the side-length 
of an initial cubic sub-volume; in the limit where the group consists of only two particles,  $\ell_{\mathrm{hull}}$ would be equivalent to $\Delta_0$, the initial separation of a particle pair.
For the density of tracer
particles in simulations NST, HC, and MC, $\ell_{\mathrm{hull}}$ varies between
$20$ and $30$ $\eta_{\mathsf{kol}}$.  The dependence of convex hull statistics on the initial length scale and density of the particle group is examined in Appendix \ref{appendixsizeden}.

Selection of each particle group based on the initial
position of the tracer particles yields groups that contain nearly the
same number of particles, with random variation of
approximately 20\% based on the homogeneous random initialization of the
Lagrangian tracer particles.  The average number of particles in a
group, $\mathsf{n_{pch}}$, listed for simulations NST, HC, and MC in
Table~\ref{simsuma}, is between 24 and 214. We follow the $\mathsf{N}_\mathsf{hull}$ convex hulls (cf.~Table~\ref{simsuma})
of the marked particle groups for the span of the simulation. The
required calculation of the hulls for each time-step is performed
using the standard QuickHull algorithm
\citep{Barber96thequickhull,2013barber}, implemented in the function
\emph{convhulln} in the package \emph{geometry} publicly available for
R, from the R Project for Statistical Computing
\citep{ihaka1996r,cranr}.  The surface area and volume of the convex
hulls are obtained based on a Delaunay triangulation of the hull
vertices.
We stop tracking the convex hull of a group of
particles when the Lagrangian crossing time, $\mathsf{LCT}$, is
reached to avoid the possibility of numerical artifacts due to the periodicity of the simulation volume.

The initial positions of particle groups could be chosen in
regions of special interest in the flow, but in this work we restrict
ourselves to a homogeneous initial distribution of the groups.  For
each simulation the ensemble of particle groups is initially selected
to fill completely a horizontal slab. The total number of groups of
particles that we analyze using convex hulls is listed as
$\mathsf{N_{hulls}}$ in Table~\ref{simsuma}. This large number of
convex hulls is more than are required for statistical convergence of
average quantities, but allows us to capture some
statistically rare flow features.

As any pair of particles separates in a turbulent flow, the particles
move with the small-scale fluctuations of the velocity field.
The distance between the two particles increases
monotonically in time on average, but any specific pair of particles
will produce an erratic, noisy signal.  If a convex hull is defined by
a very small group of particles, then most of the particles define the surface of the convex hull.  These particles on the
surface of the convex hull are called \emph{vertices} of the convex
hull.  In the situation where most of the particles are vertices, the
convex hull, like the particle-pair distance, shrinks or grows
erratically as its component tracer particles move in the turbulent
flow.  The limit where groups contain only small numbers of particles
is of little physical interest for convex hull analysis, because
particle pairs or particle tetrahedra already provide useful
dispersion information.

In simulations NST, HC, and MC we examine the relative dynamics of larger groups of
particles.  If a particle that is a vertex of the convex hull moves
inward toward the center of the larger group of particles, it is
unlikely that it will remain a vertex because of the requirement of
convexity.  It can become an \emph{interior particle} of the convex
hull.  Other particles may continue to move away from the group, and
the convex hull will typically continue to expand smoothly.  The
particles that constitute the group of vertices of the convex hull can be
exchanged frequently.  This is a distinctive concept for the convex
hull because it provides a contrast with more common Lagrangian
diagnostics such as particle pairs or particle tetrahedra. For
statistics constructed from particle pairs or particle tetrahedra, the
same particles define the size at each point in time.

The convex hull also intrinsically links a macroscopic length scale,
the size of the convex hull, with the position of the convex hull's
geometrical center.  Over this length scale, the convex hull filters
out  tracer particles which disperse slower than its vertices, selecting the
most efficiently dispersing members of the particle group. 

\subsection{Convex hull description of a group of tracer particles}

A convex hull is defined by its vertices; these are the particles that
dispersed the fastest in a given group of particles.  Potentially this
could decouple the convex hull from the enclosed
particles in two ways.  The number of vertices of the convex
hull could become extremely small, or the majority of interior
particles could detach from the convex hull vertices and clump
somewhere in a subregion inside the hull.  In this section we devise
simple basic checks for these two {scenarios}.  

\begin{figure}
 \resizebox{3.375in}{!}{\includegraphics[angle=90,width=0.5\textwidth]{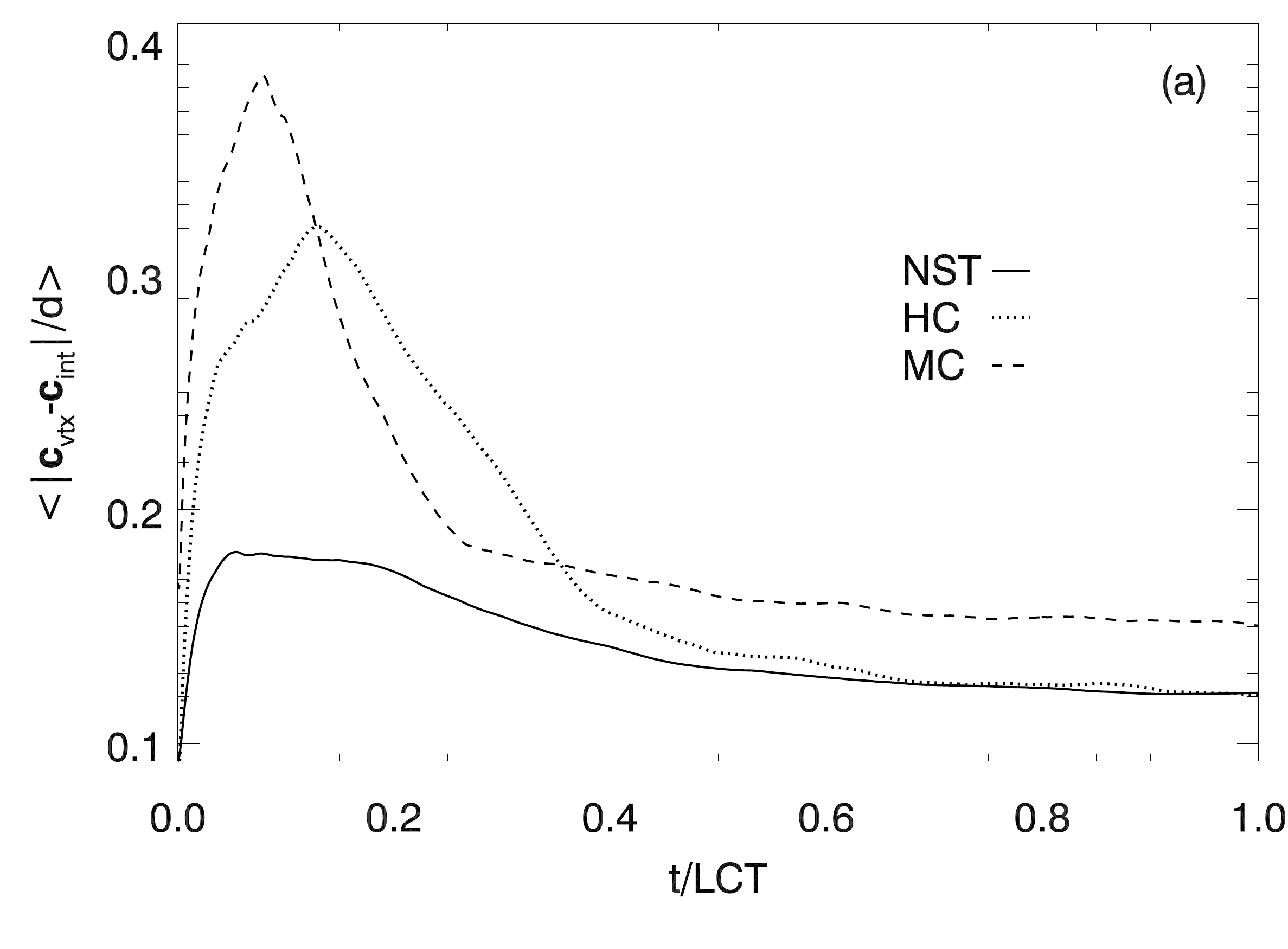}}\resizebox{3.375in}{!}{\includegraphics[angle=90,width=0.5\textwidth]{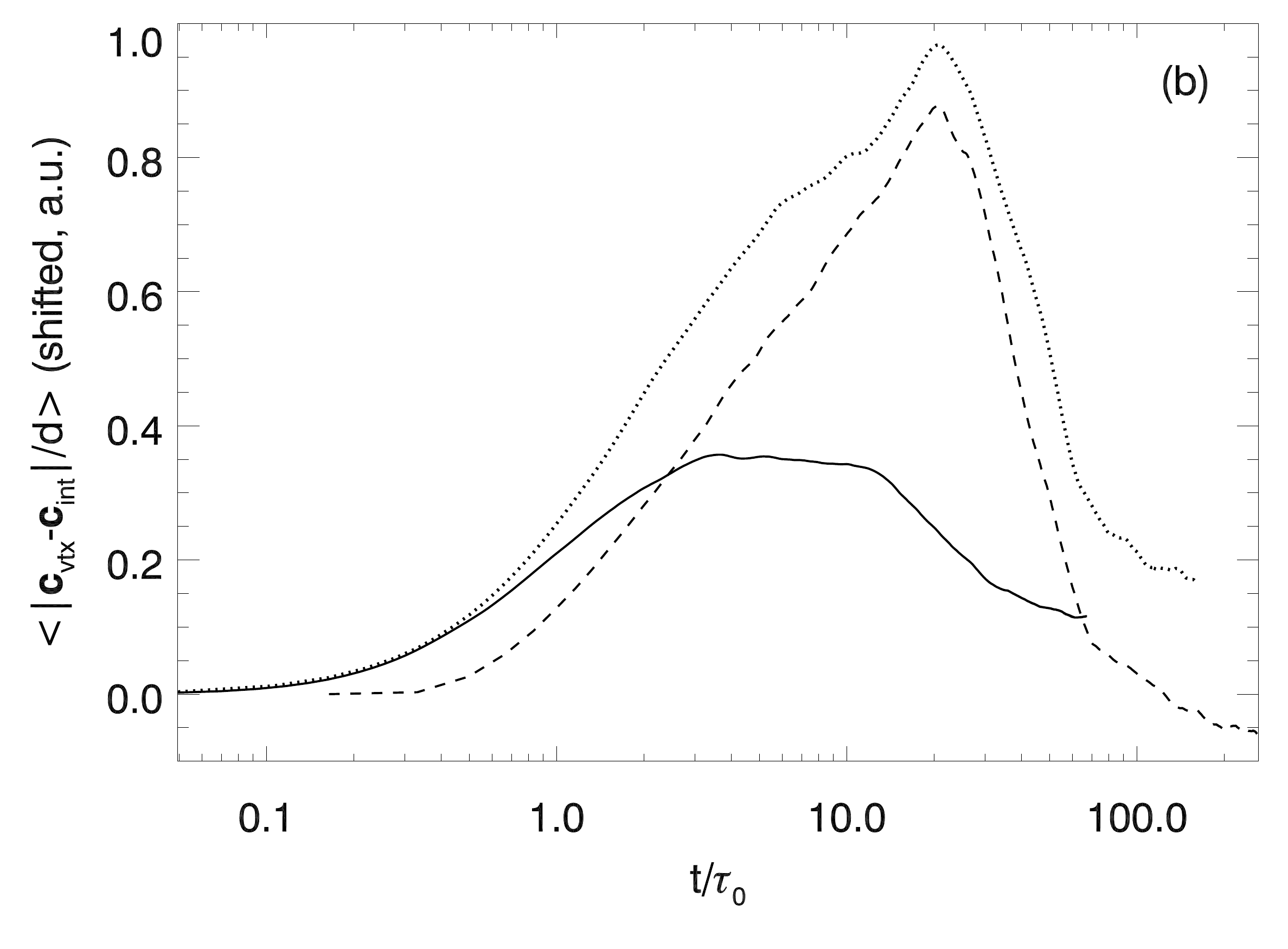}}
\caption{ 
(a) The average distance between the geometric centers of the convex hulls, $\vec{c}_\text{vtx}$, and the virtual center of mass of their interior particles, $\vec{c}_\text{int}$, divided by the convex hull size, $d=(d_{x}^2+d_{y}^2+d_{z}^2)^{1/2}$. (b) Data as shown in panel (a), {shifted vertically to common initial value}.  Solid line: NST, dotted line: HC, dashed line: MC.  
Averaging is performed over convex hulls calculated for each group of Lagrangian tracer particles and at each time.  Time is given in units of (a) the Lagrangian crossing time, LCT, and (b) the initial turnover time, $\tau_0$.
\label{newclumpgraph1} }
\end{figure}

If the particles contained in the convex hull do not spread throughout the
 space inside of the convex hull evenly as it grows, the convex hull
 will fail to characterize the full group of particles.  We use the average
 difference between the geometric center of the convex hull, $\vec{c}_{\mathsf{vtx}}$, and the
 virtual center of mass of the interior particles contained in the
 convex hull, $\vec{c}_{\mathsf{int}}$, as an indicator of decoupling. 
This difference will not be zero, because the particles that make up the
 convex hull will never fill the space perfectly evenly.  
Since this difference will grow in time as the particles disperse, we compare it to a maximal extent
 of the convex hull at any point in time, defined by $d=\left(d_x^2+d_y^2+d_z^2 \right)^{1/2}$ where $d_x$ is
the extent of the convex hull projected on the $x$-direction, and $d_y$ and $d_z$ are defined similarly.
 FIG.~\ref{newclumpgraph1}(a) shows that the average
 difference between the centers normalized by the convex hull's maximal extent, 
 $\delta c=\langle|\vec{c}_{\mathsf{vtx}}-\vec{c}_{\mathsf{int}}|/d\rangle$.  This normalized average difference between centers does 
 not become larger than 40\% during an initial phase ($0.2$ LCT $\lesssim t\lesssim 0.4$ LCT$)$  and during the subsequent 
 phase converges toward a quasi-constant level
 ranging between 15\% and 20\%, which is less than the standard
 deviation of the coordinates of the group of tracer particles for
 each simulation.  In FIG.~\ref{newclumpgraph1}(a), time is given in terms of the Lagrangian
 crossing time, LCT.

The differences in
the initial separation of the particle pairs in
FIG. \ref{pairdispfig} ($\Delta_0\simeq \eta_{\mathsf{kol}}$) and the mean initial
length scale of the convex hulls ($\ell_\mathrm{hull}\simeq 20-30 \eta_{\mathsf{kol}}$)
generate dispersion curves that reflect different ranges of temporal and spatial
scales of the underlying turbulence. Because the
observable dispersion regimes and their duration can change as a
consequence of different $\Delta_0$ or $\ell_\mathrm{hull}$, 
a direct comparison of both figures is difficult. In Navier-Stokes turbulence, the initial turnover
time, $\tau_0$, has been shown\cite{bourgoin:disppheno,bitane_homann_bec:disptimescale} to signal the transition from the
ballistic to the inertial range of dispersion, 
and thus to provide a reference scale of dispersion. We therefore use the initial turnover
time $\tau_0$ to normalize the dispersion of particles contained in a convex hull, with initial length scale $\ell_\mathrm{hull}$. It is however not expected that
this normalization can eliminate
the physical differences between isotropic Navier-Stokes and anisotropic convective systems. Moreover,
it is not clear whether the universality of $\tau_0$ extends beyond the transition from ballistic
to Richardson-like dispersion.

Close examination of FIGs.~\ref{pairdispfig} and
\ref{newclumpgraph1} shows that the initial phase, during which the average difference in the centers of
convex hull and interior particles increases to a maximal value, extends into the fast separation regime of
particle pair dispersion. 
The subsequent phase of decreasing $\delta c$ corresponds to separation scales near to and in the diffusive regime. These signatures, as well as the sharp transients
evident between phases of the evolution of $\delta c$ in FIG. \ref{newclumpgraph1}, indicate a potential utility of the convex hull for studies of the turbulent inertial range.

\begin{figure}
\resizebox{3.375in}{!}{\includegraphics[angle=90]{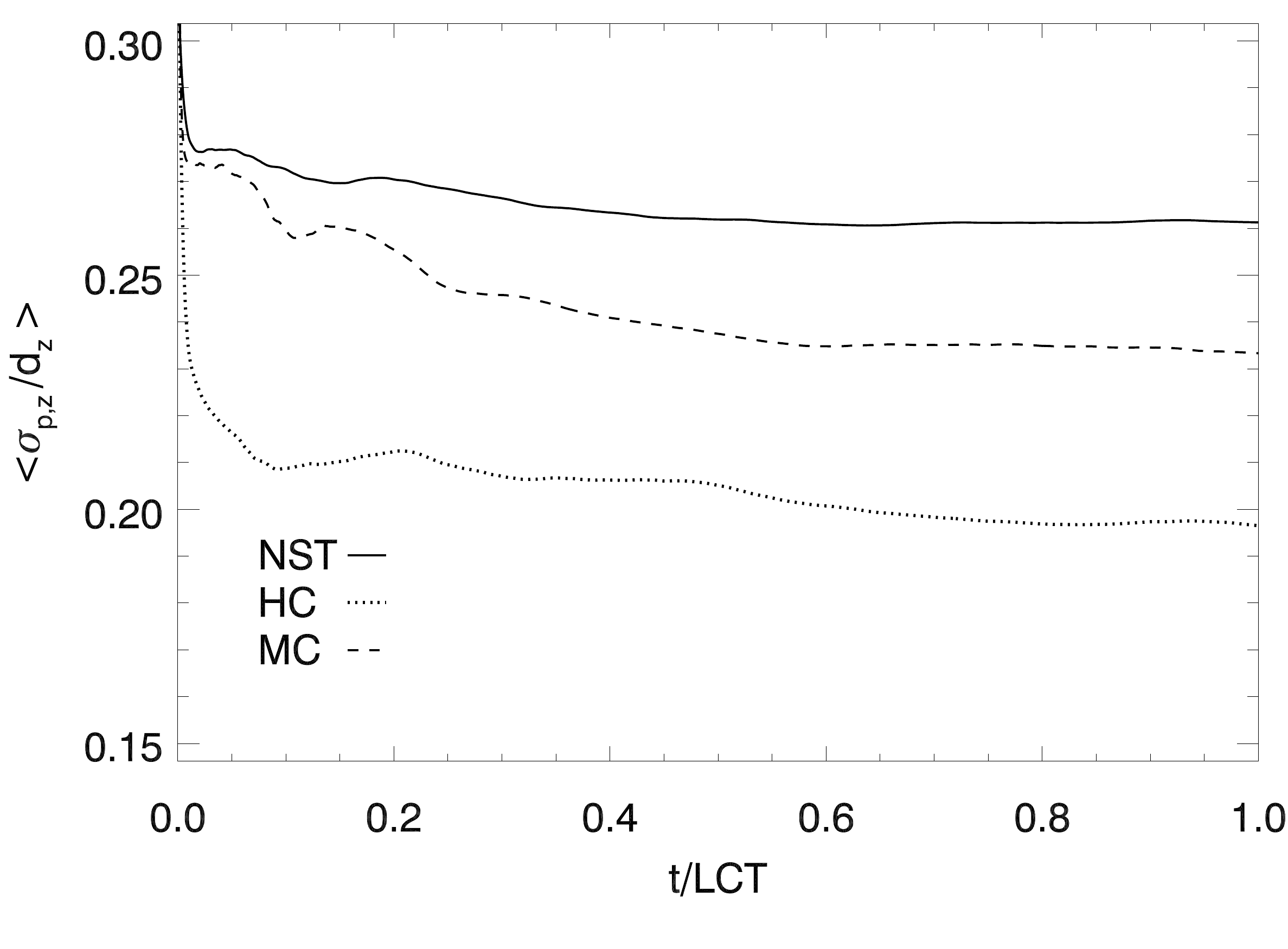}}
\caption{ 
The standard deviation $\sigma_\text{p,z}$ of the $z$ coordinates of interior particles of a convex hull
divided by its extension along the z-direction $d_{z}$. Averaging is performed over all convex hulls in each simulation.  Time is given in units of the Lagrangian crossing time, LCT. 
\label{newclumpgraph2} }
\end{figure}

FIG.~\ref{newclumpgraph2} reveals the 
distribution of the group of particles within the convex hull in the $z$-direction.  Here the
$z$-direction has been selected because it is the direction of the
gravitational anisotropy in the convective cases; however for one-dimensional cuts in
directions other than the $z$-direction, similar curves result.  The
ratio plotted in FIG.~\ref{newclumpgraph2} is the standard deviation
of the particle positions, $\sigma_\text{p,z}$, divided by the extent of the hull in the $z$-direction.  This
ratio would be small if many of the tracer particles were to form a
clump rather than spreading throughout the interior of the convex
hull.  For each of the three simulations we study, however, this
quantity quickly comes to a plateau.  After $0.1$ to $0.2$ LCT, i.e. the scales probed by the particles
as they approach the inertial range, the
ratio no longer decreases substantially.

In all simulations the average number of vertices of the convex hulls
decreases only mildly with time; this decrease is on the order of 10\%
before the Lagrangian crossing time is reached.  After the short
initial phase up to $\tau_0$, the decrease in the number of
convex hull vertices happens very gradually.

We conclude that on average in simulations NST, HC, and MC, the convex hulls and their interior particles  
do not detach from each other 
in a way that would render the concept of the
convex hull inappropriate for characterizing a
pre-selected group of many Lagrangian particles.  
Based on the measurements presented,  a clear distinction can be made 
between the diffusive regime and the inertial range. The correlated inertial-range velocity fluctuations
lead to changes in the relationship of convex hull vertices and interior particles. This trend is reversed 
as soon as the diffusive regime is reached, largely neutralizing the differences between interior particles
and the convex hull vertices on 
inertial scales. This susceptibility of the convex hull to the different characteristic regimes
of ballistic, inertial-range, and diffusive turbulent transport render this diagnostic 
attractive for future Lagrangian investigations of turbulence.

Apart from the ability of the convex hull to indicate different regimes of turbulent transport,  
the tests above also yield information
about the dynamics of the turbulent velocity field.
The average displacement shown in FIG.~\ref{newclumpgraph1} quantifies anisotropic  
differences between the dynamics of the most efficiently dispersing convex hull vertices 
and the slower dispersing interior particles. On the spatial scale
set by the convex hull, an 
anisotropic difference of the velocity fluctuations responsible for vertex and interior tracer transport
is observable as
a relative displacement of the centers of the group of interior particles and of those 
that define the convex hull.
FIG. \ref{newclumpgraph1}(b) which is a different representation of the data shown in FIG. \ref{newclumpgraph1}(a)
demonstrates this point.
All three systems have slightly different initial Lagrangian tracer configurations and, consequently,
the corresponding initial values of $\delta c$ differ by up to $8$\% (MC) while NS and HC have an initial
difference of approximately $1$\%. { Shifting the $\delta c$-curves of all three systems to a common initial level}   
allows a qualitative comparison although this simple approach can not eliminate all dynamical 
differences caused by varying initial tracer separations.

The increase observed for $\delta c$ is driven by the particles that are part of the 
surface of the convex hull, since they determine the geometric center of the convex 
hull. The relative motion of particles contained in the interior of the hull is driven by velocity fluctuations on scales smaller than the convex hull size. 
Particles at opposite locations on the surface of the convex hull will experience velocity differences
on the scale of the convex hull and therefore tend to move more rapidly apart from each other
than particles in the interior of the hull, which in turn determine the center of mass of the convex 
hull. Thus on time scales of $(t-t_0)\lessapprox\tau_0$ a significant displacement between the geometric 
center and the center of mass of a group of particles can occur, evidenced in the rapid 
growth of $\delta c$. The relative displacement of the geometric center and the center of mass
continues to grow at a slower rate for $(t-t_0)> \tau_0$.
This can be attributed to a finite time correlation of the velocity fluctuations on the scale of the convex hull. 
In addition, as time evolves and the hull grows in size, particles in the interior of the convex hull will also experience increasing 
velocity fluctuations and thus some interior particles may become particles on the surface of the hull and 
- vice versa - particles on the surface of the convex hull can move into its interior due to engulfment by other particles.
This process eventually leads to a decrease in the relative displacement of the geometric center and the center of mass as the
diffusive regime is approached. A noticeble difference between the NST configuration and the convective systems HC and MC is the presence of a plateau for
the NST case between $3\tau_0$ and $16\tau_0$, while for HC and MC, $\delta c$ continues to grow during this time.
The different behavior may be caused by anisotropy in the convective flows HC and MC sustaining longer correlations in time for
velocity fluctuations in preferential directions, which does not occur for the statistically isotropic Navier-Stokes case.

The quantity shown in FIG.~\ref{newclumpgraph2} measures  
the diffusive character of the motion of the interior particles, rather than dynamical anisotropy.
This measure exhibits a rapid transient around $\tau_0$ 
from initial levels towards a first roughly  constant plateau throughout the inertial range
that finally approaches the asymptotic diffusive value around $(t-t_0) \simeq 100\tau_0$.   
Here, the inherently hydrodynamic simulations NST and HC display less variation throughout the inertial 
range than system MC which exhibits additional flow structuring due to the presence of magnetic field
fluctuations.
This brief interpretation allows for extensions,
for example focusing on vertex dynamics or a  
detailed direction-specific analysis by introducing spatial projections of the hulls to narrow 
down the structure of the underlying anisotropic fluctuations.  This will be subject of future work.

\subsection{Multi-particle dispersion using convex hull analysis \label{secmaxray}}

Because ballistic and diffusive ranges for 
particle pair dispersion are typically discussed in terms of
length squared, we employ analogous measures for a group of particles
and convex hulls.  This is intended to make comparison with
dispersion curves as simple and direct as possible.  We calculate a
maximal ray $r$ internal to a convex hull defined by a group of
particles $G$:
\begin{eqnarray}\label{maxraydef}
r = \max_{i,j \in G} \sqrt{(x_i - x_j)^2+(y_i - y_j)^2+(z_i - z_j)^2}
\end{eqnarray}
By definition, the particles $i,j$ that contribute to the maximum in
this definition are always vertices of the convex hull.  If the group
of particles densely filled a sphere, the convex hull would be the
surface of the sphere, and the maximal ray would be the
diameter of the sphere.  
For this reason the maximal ray is
sometimes also called the diameter of a convex hull. However in this
work we examine anisotropic systems where the convex hull of a group
of particles is not typically close to spherical; we opt for the more
accurate former term.
The susceptibility of the maximal ray's orientation to deformations of the 
convex hull can be parameterized by the 
RMS value of the vertex distance from the hull's geometrical center normalized by the average distance to the center (averages
taken over the hull vertices),
$Q=\sigma_{r_\text{vtx}}/\overline{r_\text{vtx}}$. 
If $Q \approx 0 $, i.e. the convex hull is close to spherical,  
the maximal ray can change its direction by an arbitrary amount and much faster 
than the autocorrelation times of the underlying turbulent fluctuations would suggest. 
In this case, small fluctuations of the hull radius, which can occur due to uncorrelated small-scale
fluctuations, will lead to rapid changes of orientation of the maximal ray.
The maximal ray is highly susceptible 
to anisotropic deformations of the convex hull.
In contrast, a significant anisotropic deformation of the convex hull, $Q\gg 1$, acts like a threshold for 
directional variation of the maximal ray and stabilizes its orientation.
This subsection focuses on 
quantities specific for the convex hull and their relation to classical Lagrangian mean-square 
pair-separation, $\langle(\Delta-\Delta_0)^2\rangle$.

\begin{figure}[H]
 \resizebox{3.5in}{!}{\includegraphics[angle=90]{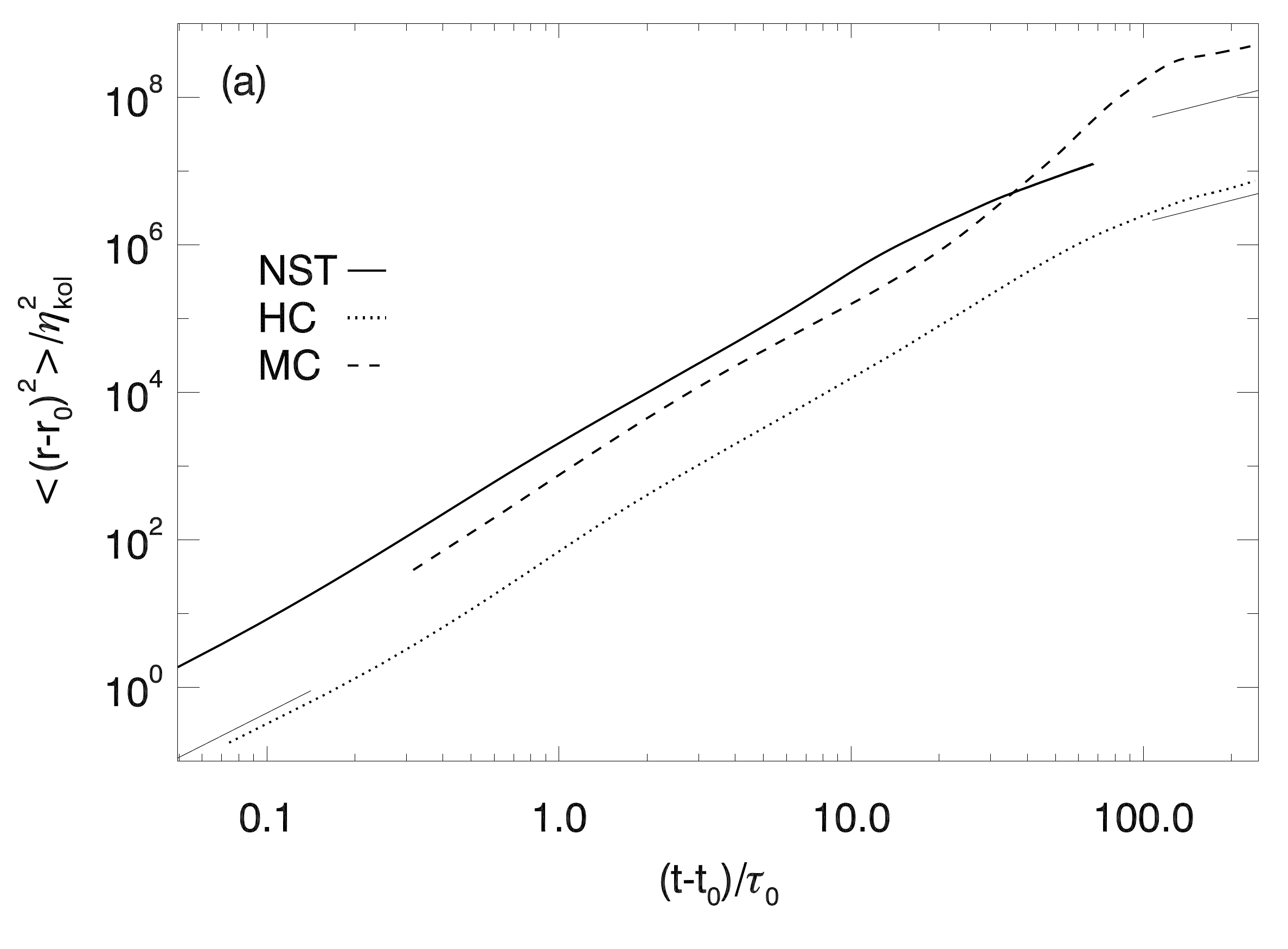}}
 \resizebox{3.5in}{!}{\includegraphics[angle=90]{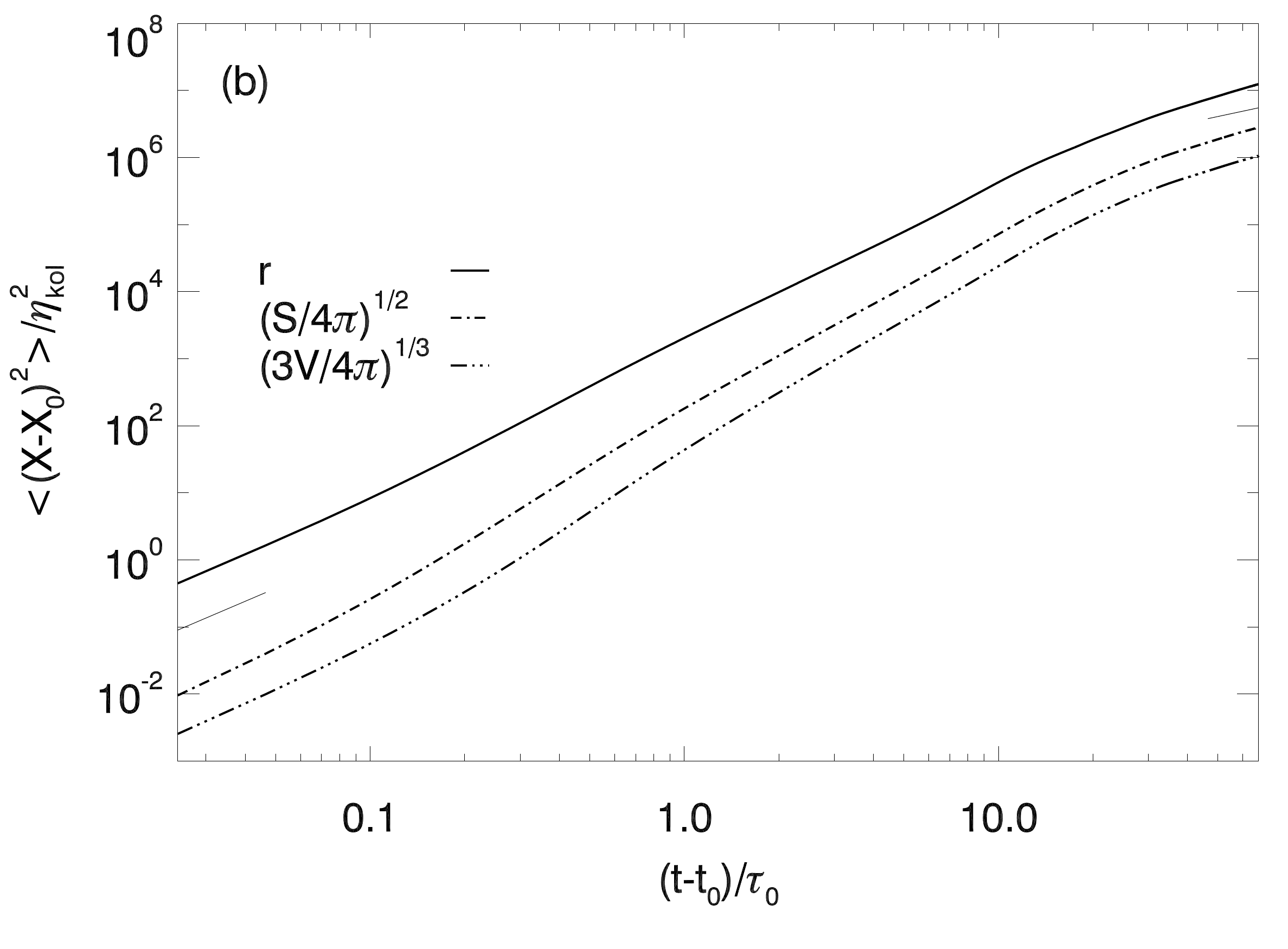}}
\caption{(a) Evolution of the mean-square maximal ray $r$ of the convex hulls in all three systems. (b)~Evolution of mean-square maximal ray $X=r$ (solid curve), of the length based on the hull's surface area, $X=(S/4\pi)^{1/2}$ (dot-dash), and of the length based on the hull's volume, $X=(3V/4\pi)^{1/3}$ (dash-3dot), for simulation NST, thin solid lines as in FIG.~\ref{pairdispfig}. Brackets indicate averaging over all groups of tracer particles in a horizontal slab in each simulation volume. The symbols $r_0$, $S_0$, and $V_0$ denote the respective initial values.\label{dispgraph} } 
\end{figure}

We average the square of the maximal ray over all groups of particles; the
results for each simulation are shown in 
FIG.~\ref{dispgraph}(a).  This figure demonstrates
that the maximal ray, although it is not tied to the same particle pair in
each tracer group, asymptotically converges to a ballistic regime
signature $\sim t^2$ up to approximately $\tau_0$, and an asymptotic
diffusive regime $\sim t$ at long times, for all 
systems considered. The data shown for MC does not attain the same
temporal resolution as for systems NST and MC due to a larger 
time step but penetrates further into the diffusive
regime.

Additional length-scale estimates can be obtained from taking appropriate powers of the 
normalized surface area, $r_\text{S}=(S/(4\pi))^{1/2}$, and the volume, $r_\text{V}=(3V/(4\pi))^{1/3}$, of the convex hulls.
Averaging the length scales
produced by many different particle groups reveals dispersive
behaviors that also tend to obey the ballistic and diffusive scaling
laws.  A comparison of dispersion curves produced from the surface
 area and volume of simulation NST are shown in 
FIG.~\ref{dispgraph}(b).  Similar to 
 Lagrangian pair dispersion, 
 the expected asymptotic 
 scaling laws for ballistic and diffusive regimes are approached by the surface- and volume-based distance
approximation. However, they hold over a shorter period
 of time than those shown in FIG. \ref{pairdispfig}.
 Although FIG.~\ref{dispgraph}(b) shows dispersion curves only for simulation NST,
 similar results are found for simulations HC and MC.  A
 Richardson-Obukhov-like regime is not observed.  
 Because achieving a clear Richardson-Obukhov regime in direct numerical simulations depends on the initial separation of particles as well as the size of the inertial range, 
 a Richardson-Obukhov regime is not expected in our simulations.  Particle filtering, an inherent property of the selection criterion of convex hull vertices, may also contribute to the lack
 of a clear Richardson-Obukhov regime resulting from convex hull analysis of dispersion. 
 During early dispersion, the vertices of a convex hull tend to be particles that move away from the center of the hull most rapidly in the direction radially outward from the center
 of the particle group;  this may explain the quasi-ballistic signature
 before approximately $16 \tau_0$ (cf. FIG. \ref{newclumpgraph1}(b)).
 As noted by
 \citet{bianchi2016evolution}, although there is a conceptual
 connection between many-particle groups and particle pairs,
 many-particle groups provide different information when measuring
 dispersion scalings.

There  is a fundamental difference between the maximal ray and the surface- or 
volume-based length approximations that becomes particularly important with regard to  
deformations of the convex hull: the maximal ray by definition runs along the direction of 
maximum extent of the convex hull. In contrast, 
the other two quantities yield averaged and isotropized approximations of the length scale probed 
by the hull, i.e. the radius of a reference 
sphere of same surface or volume. Spherical geometry is a natural first-order approximation
of a convex hull or, more precisely, the convex polyhedron, which we use as its numerical representation, since
convexity implies that the hull has no corner pointing inwards. This constraint severely restricts 
the complexity of the hull's surface structure, since any such corner vertex would turn into
an interior point enclosed by the hull. This results in an object which can mainly be deformed by
flattening of the inscribed spheroid along some direction perpendicular to the maximal ray. 
The convex hull is not material and therefore is not constrained by volume conservation in incompressible
flow.    
Although the possible length definitions do not show large qualitative differences compared to the maximal ray,
their behaviour relative to each other reflects the different responses of hull area and volume
to deformations of the convex hull. This will be exploited in Section \ref{section5}.
         
\begin{figure}
\resizebox{3.375in}{!}{\includegraphics[angle=90]{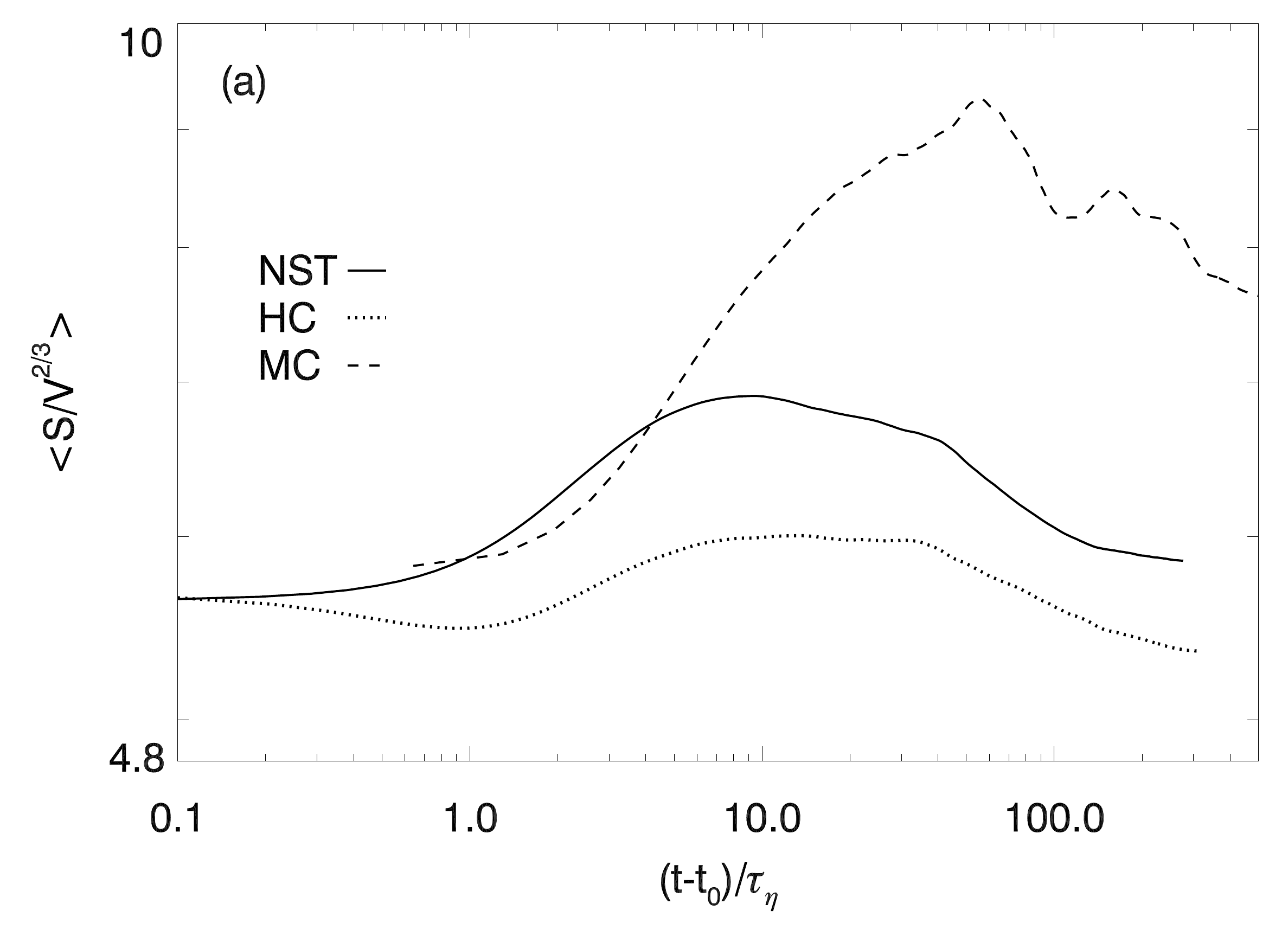}}\resizebox{3.375in}{!}{\includegraphics[angle=90]{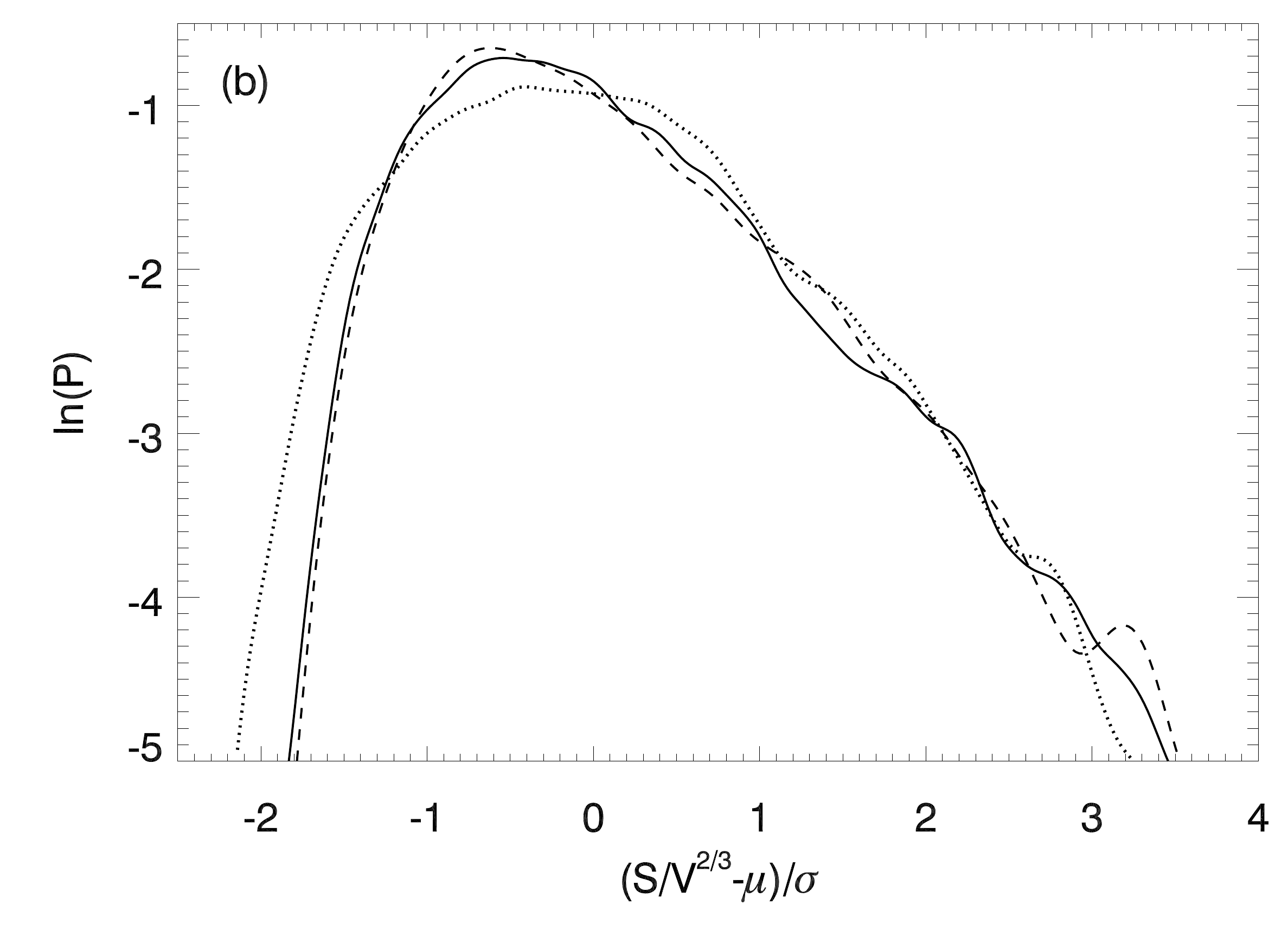}}
\caption{The time evolution of the convex hull's surface area, $S$, divided by the $2/3$ root of the volume, $V$.  In (a) the evolution of this non-dimensional surface-volume ratio is averaged over all convex hulls in each simulation.  In (b) the probability distribution function, $P$, of $(S/V^{2/3}-\mu)/\sigma$ is shown at time 20 $\tau_\eta$, $\mu$ denoting the mean and $\sigma$ the standard deviation of the respective distribution. 
The tuples $(\mu,\sigma)$ are NST:~(6.8,0.7), HC:~(6.0,0.3), MC:~(8.5,1.6).     
\label{hullvol} }
\end{figure}

{
\section{Results: Anisotropic dynamics of convex hull vertices \label{section5}}

The relationship between the surface area, $S$, and volume, $V$, of a convex hull
reveals the anisotropy of vertex transport in a turbulent flow, which is of
particular interest during convection, i.e. in the presence of coherent velocity structures.  We introduce the
non-dimensional ratio $S/V^{2/3}$ as a direct way to quantify
anisotropy.  Because a sphere minimizes the amount of surface area for
a given volume, an absolute lower bound of $4\pi / (4/3
\pi)^{2/3}\approx 4.8$ exists for this non-dimensional surface-volume
ratio.  An
anisotropic convex hull, e.g. a cigar-shaped or a pancake-shaped hull, will
have a higher surface to volume ratio, so the ratio gives an
impression of how non-spherical the current state of the hull is. The
ratio cannot differentiate between prolate (cigar-shaped) and oblate
(pancake-shaped) convex hulls, because it approaches infinity in the limit
both of zero pancake thickness and infinite cigar length.  Higher values indicate a basic level of anisotropic deformation. 
FIG.~\ref{hullvol}(a) shows the time evolution of the surface-volume
ratio, averaged over all convex hulls in each simulation.  Because the
particle groups consist of small numbers of particles which are
randomly distributed, they are not initially perfectly isotropic and
do not evenly fill the cubic initial volumes; the resulting convex
hulls do not form either perfect cubes or perfect spheres. Thus the
surface-volume ratio initially exhibits an average value of
approximately 5.6, a low value that lies between the values for
perfectly spherical and perfectly cubical volumes.
In all
simulations, the surface-volume ratio begins to increase around $t=\tau_\eta$,
indicating that the convex hulls typically become stretched,
\emph{i.e.} anisotropic, as their particles start to disperse due to turbulent fluctuations.
 In the Navier-Stokes case (NST) no global anisotropy exists
in the flow. As expected, the average surface-volume ratio remains
relatively low throughout the simulation reaching its maximal value around 10 $\tau_\eta$.
At long times, the average
surface-volume ratio returns to approximately its initial value as uncorrelated particle motion begins to
eliminate anisotropic deformations of the convex hull.  The
changes in the surface-volume ratio also slow and it approaches a flat
regime related to the diffusive trend observed in FIG.~\ref{dispgraph}
at long times.  

In the case of hydrodynamic Boussinesq convection (HC), the mean
 temperature gradient introduces a preferential direction. We would
 thus straightforwardly expect higher stretching of the hulls in this
 direction. However, FIG.~\ref{hullvol} shows that this does not take
 place for the convex hulls we followed; for times greater than $\tau_\eta$ only a slight
 increase occurs followed by a plateau phase up to $30 \tau_\eta$. Subsequently,
 the average surface-volume ratio quickly
 decreases below its initial value.  The scale of the convective
 plumes in simulation HC are large and diffuse, as reflected by the
 large Bolgiano-Obukhov length $\ell_{\mathrm{bo}}$, the smallest
 scale on which the cascade of thermal fluctuations is driven by
 buoyancy \citep{biskampbook}.  This large Bolgiano-Obukhov length
 indicates that smaller-scale turbulent dynamics are not driven by the
 anisotropic influence of buoyancy.  The convex hulls do not tend to
 become strongly anisotropic, because the length scale of the
 anisotropic convection differs considerably from the scale of the
 convex hulls examined. The Reynolds number of HC which is less by approximately $50\%$ as compared
 to the value of the NST system also explains why the HC simulation exhibits the lowest 
 level of convex-hull anisotropy.  
 
 A different behavior is observed for the
 magnetohydrodynamic convection (MC) simulation since larger-scale magnetic fluctuations
 have a strong impact on small-scale dynamics (in contrast to a large-scale velocity there exists 
 no frame of reference which eliminates
 the magnetic field); consequently, far higher
 surface-volume ratios are attained than in the other two cases. In
 this simulation the large-scale magnetic field fluctuations result in
 strong local anisotropy of the small-scale velocity fluctuations
 \citep{grappin2010scaling,verdini2015anisotropy,matthaeus1996anisotropic,cho2000anisotropy,chandran2008strong,montgomery1981anisotropic,goldreich_sridhar:gs2,boldyrev:bmodelII};
 the consequence is considerable stretching of the convex hulls.

The mean alone does not characterize the full information that the
convex hull analysis can provide about anisotropy in each
simulation.  
The shape of the probability distribution of the
surface-volume ratio yields a more comprehensive picture.
If all convex hulls in a simulation were perfect spheres,
the distribution of the surface-volume ratio would be a delta
function. 
However,  the distributions show a strong dependence on the type of
turbulence indicated by 
the values of distribution mean, $\mu$, and standard deviation, $\sigma$, given in the caption
of FIG. \ref{hullvol}. 
In the hydrodynamic convection case the distribution is
the narrowest with the lowest mean indicating the highest level of anisotropy, followed
by NST and MC.  
The significant hull anisotropy observed for system MC is a clear indication of the additional
anisotropy imposed by the slowly evolving large-scale magnetic field
fluctuations on the smaller-scale velocity fluctuations.
These results are not surprising and consistent with the 
data given in FIG.~\ref{hullvol}(a). In addition, FIG.~\ref{hullvol}(b) shows the
centered and normalized distributions of the surface-volume ratio for each of our three
simulations after each set of convex hulls has evolved for 20
$\tau_\eta$. All distributions collapse on a positively skewed functional shape suggesting a general 
characteristic of convex hull deformation common to all three turbulent systems.    
}

\begin{figure}
\resizebox{3.375in}{3.00in}{\includegraphics[angle=90]{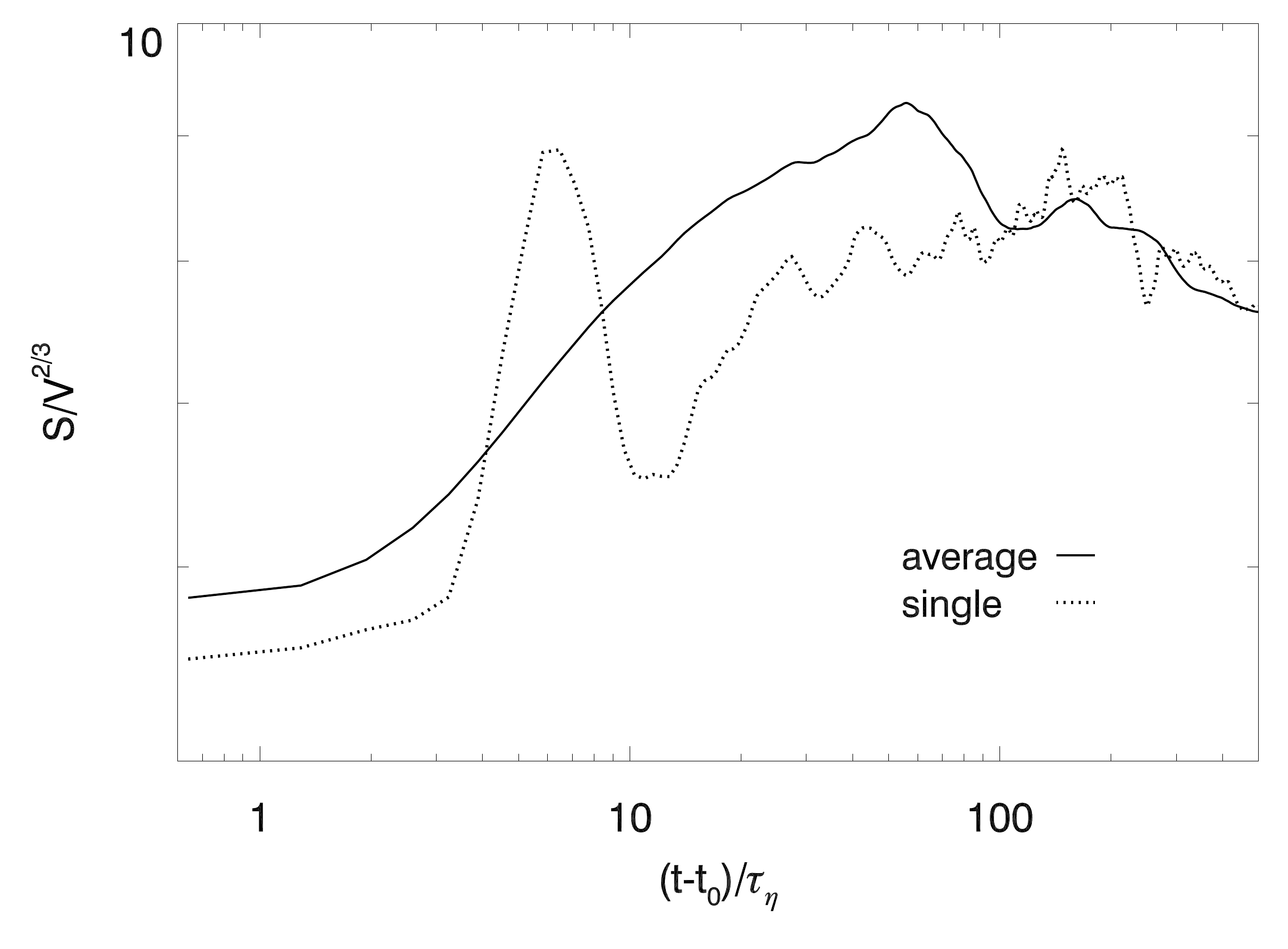}}\resizebox{3.375in}{!}{\includegraphics{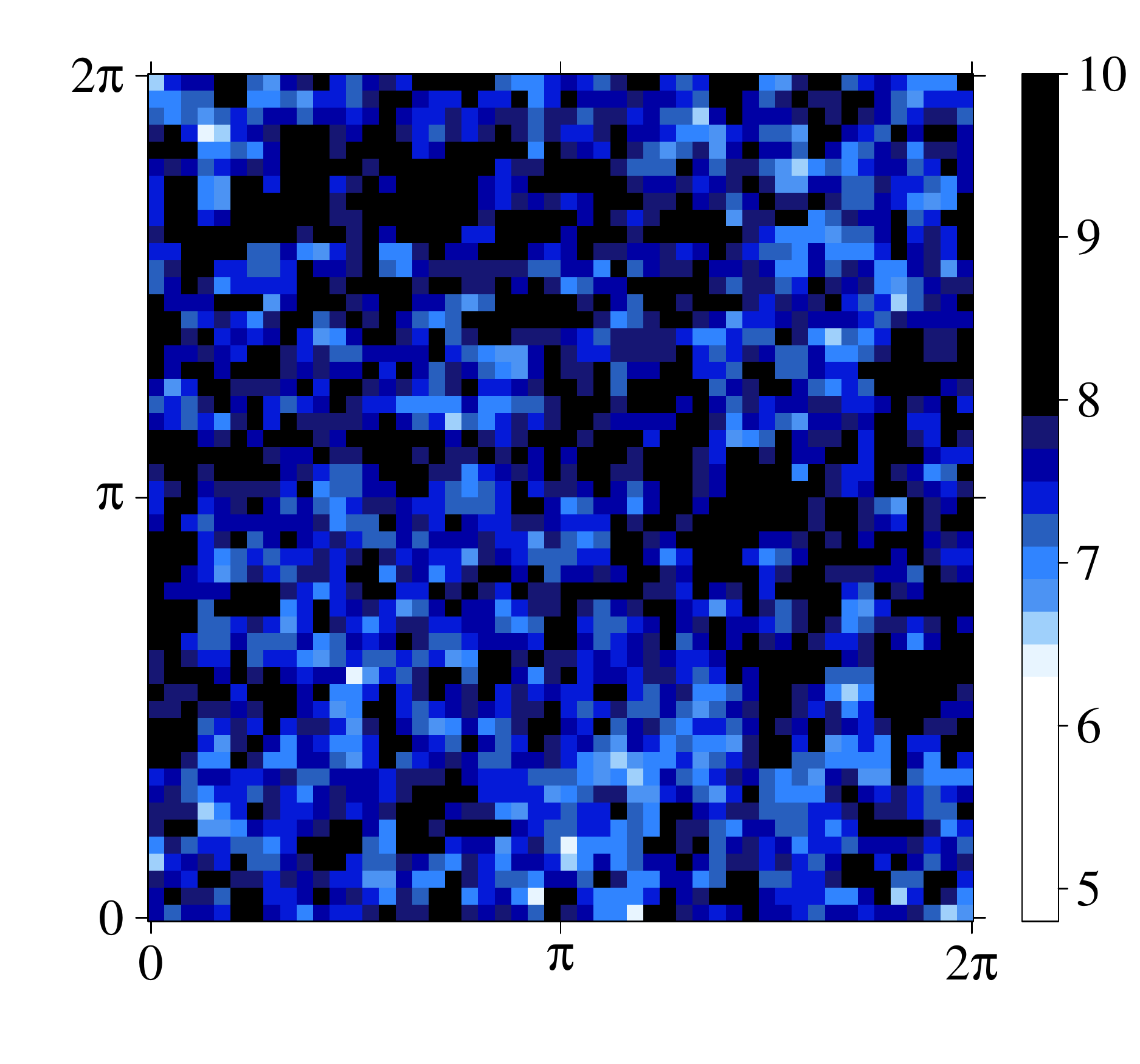}}
\caption{
(Left) a comparison of the non-dimensional surface-volume ratio between the convex hull of a single arbitrarily chosen group of tracer particles and the average,  in the simulation MC.  
(Right) a contour plot that shows a horizontal slab filled with convex hulls in simulation MC, at a late time in the simulation.  Darker colors represent higher values of the surface-volume ratio.  The colors are shown at the initial positions of the convex hulls, and each pixel approximately represents the initial volume of a convex hull. 
\label{anipdf} }
\end{figure}

The surface-volume ratio varies spatially in each simulation.  The time evolution of this ratio for a single convex hull in simulation MC  is illustrated in FIG.~\ref{anipdf}.
At early times, the surface-volume ratio for this individual hull grows to considerably exceed the mean, indicating that this hull is more stretched than the average convex hull of this ensemble. This surface-volume ratio also exhibits rapid changes in time.  For example, during the period between approximately 5 $\tau_\eta$ and 10 $\tau_\eta$ this hull goes from a more anisotropic form than average to a considerably less anisotropic form.

In FIG.~\ref{anipdf}, the surface-volume ratio is also shown as a contour plot for the set of convex hulls that fill a horizontal slab of simulation MC. Dark areas represent regions where convex hulls have grown with significant anisotropy. High spatial intermittency is also noticeable, with areas of large anisotropy bordering areas that grow more isotropically. This pattern of anisotropy remains similar over a long period of time, reflecting the strong influence of the initial configuration of the flow on local dispersion.
Although we examine a small number of simulations, the non-dimensional surface-volume ratio that we introduce is clearly capable of revealing aspects of local anisotropy in turbulent flows.

\section{Results: Extreme-value statistics of turbulent particle dispersion \label{section6}}

The vertices of a convex hull are the particles that disperse fastest among a given group of particles, and the maximal ray defines a maximal dispersion of all particle pairs within the group.  Thus the use of the convex hull evokes concepts from extreme value theory \citep{castillo2005extreme,majumdar2010random}. The most widely encountered distribution in extreme value theory, the Gumbel distribution\citep{bramwell2000universal,gumbel1958statistics}, has been frequently employed for climate modeling, including extreme rainfall and flooding \citep{hirabayashi2013global,borga2005regional, koutsoyiannis2004statistics,coles2003fully,yue2000gumbel}, extreme winds \citep{kang2015determination}, avalanches \citep{schweizer2009forecasting}, and earthquakes \citep{pisarenko2014characterization}. The Gumbel distribution has also been found to reasonably characterize the density fluctuations within galaxies \citep{antal2009galaxy,waizmann2012application,chongchitnan2012primordial} and in certain areas of tokamaks \citep{hnat2008characterization,anderson2009predicting,graves2005self}, binding energies in liquids \citep{chempath2010distributions}, as well as turbulent fluctuations \citep{noullez2002global,dahlstedt2001universal}. 
   The cumulative distribution function $F$ for the Gumbel case has the well-known form:
\begin{eqnarray}\label{eqgumbel}
F(x)=\exp{(-\exp{(-(x-\mu)/\beta)})}
\end{eqnarray}
where the location parameter $\mu$ gives the mode of the distribution, $\beta$ is commonly called the scale parameter, and the median of the distribution is $\mu-\beta \ln(\ln(2))$.   Because extreme value theory typically develops as an asymptotic theory for sample sizes $n \sim \infty$,  convex hulls with large numbers $n$ of particles facilitate the exploitation of extreme value theory results.

We examine the square-length of the maximal ray with extreme value theory, and this choice is crucial.  The square-length of the maximal ray is a fundamental scalar commonly associated with dispersion, and thus the most natural physical quantity to consider.  The square-length of the maximal ray is also consistent with a simple model of Gaussian displacements.  No rigid upper limit exists for the square-length of the maximal ray, and thus the Gumbel distribution is the case that would be anticipated from extreme value theory.  

Because Lagrangian tracer particles move in a flow with a finite correlation in space and time, their motions are not independent.  The number of particles in each
 group is also limited in these numerical experiments.  Despite these limitations, we find that the shape of the cumulative
 distribution function of the square of the maximal ray is suggestive of a Gumbel distribution.  This observation
 holds at each point in time, regardless of whether the particle groups sampled are in the ballistic regime, diffusive regime, or a transitional period of dispersion.  A Gumbel
 distribution describes the results well, regardless of the initial length scale of the convex hull, and the initial density of particles, for the range $4 \eta_{\mathsf{kol}}< \ell_{\mathrm{hull}} < 64 \eta_{\mathsf{kol}}$ that we have tested (see Appendix \ref{appendixsizeden}).
 This suggests that the Gumbel distribution might provide an effective description of the probability of extremes of turbulent dispersion.  The location and scale parameters can be different for different $\ell_{\mathrm{hull}}$, and at different times in the dispersion process, although a Gumbel distribution is recovered at each time.
 
\begin{figure}
\resizebox{3.375in}{!}{\includegraphics{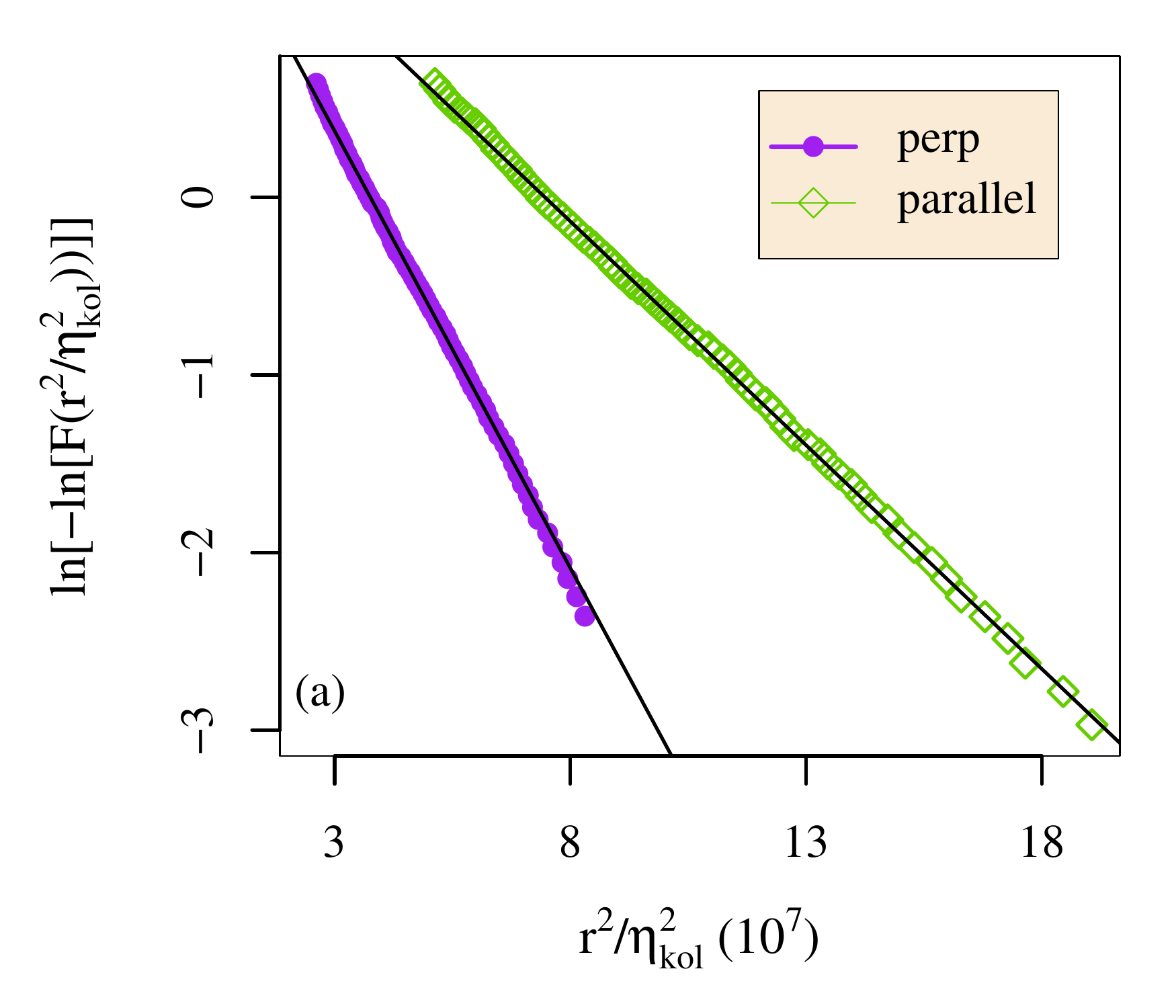}}\resizebox{3.375in}{!}{\includegraphics{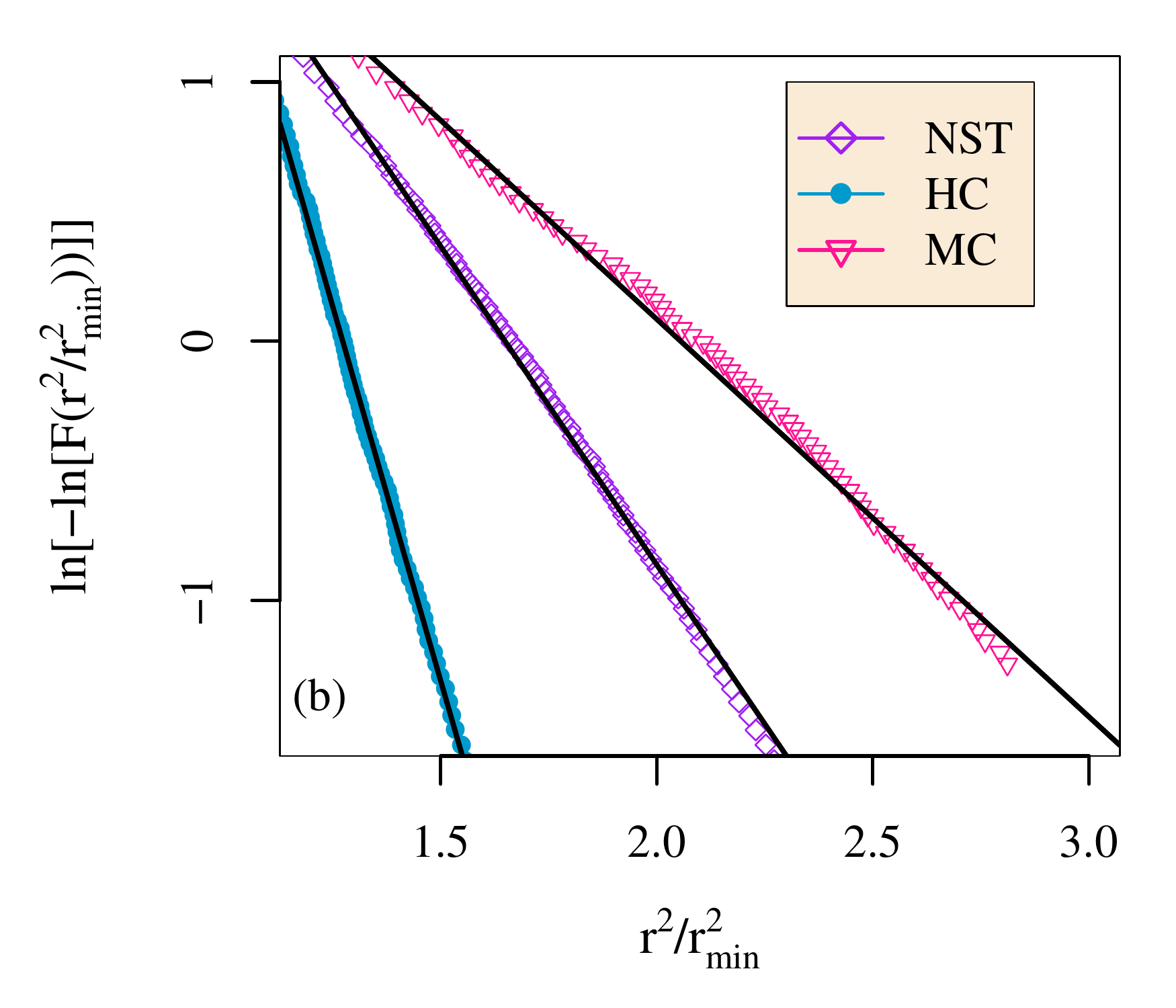}}
\caption{The log negative log of the cumulative distribution function, $F$, of the square of the maximal ray of the group of particles defined in eq.~\eqref{maxraydef}.
  Panel (a) shows the cumulative distribution function of the square of the maximal ray in the directions perpendicular and parallel to gravity from the MHD convection simulation MC. In (b) shows the cumulative distribution function of the square of the maximal ray for each simulation.
  For each cumulative distribution function shown, a line (solid black line) fits the natural log of the negative natural log of $F$ well.  
\label{extremepdf} }
\end{figure}

In addition, we consider a cumulative distribution function constructed from data at all times throughout the evolution of the convex hulls, as shown in FIG.~\ref{extremepdf}.  Using data from all times is a reasonable choice that produces a single form of the cumulative distribution function relevant to the entire simulation.
From the perspective of the simple model of Gaussian displacement, noted above, that pragmatic choice actualizes a distribution of values of the scale parameter.  Such a possibility is well known in related, but physically distinct, studies of turbulence \citep[e.g.][]{castaing1990velocity}.
  FIG.~\ref{extremepdf}(a) shows that the distribution of square-length of the maximal ray is fit well with a Gumbel distribution when physically distinct directions, perpendicular and parallel to gravity, are considered individually in the magnetoconvection simulation MC.  
    We found in Section \ref{section5} that the convex hulls in simulation MC become highly anisotropic on average.  Thus the fact that a Gumbel distribution with different location and scale parameters accurately describes the extremes of dispersion in both of these physically distinct directions is a new and significant physical observation.

In FIG.~\ref{extremepdf}(a), we observe an ordering between the scale parameter obtained for the direction perpendicular to gravity and the direction parallel to gravity; the value of the scale parameter is larger in the direction parallel to gravity.
 FIG.~\ref{extremepdf}(b) compares the cumulative distribution functions of the square-length of the full maximal ray of the convex hull in each simulation, and again they demonstrate the linear behavior expected of $\ln{(-\ln(F))}$ for the Gumbel distribution.   When the $\ln{(-\ln(F))}$ is fit using linear regression, the value of the scale parameters are: $\beta_{\mathsf{HC}}=0.17$,  $\beta_{\mathsf{NST}}=0.40$, and $\beta_{\mathsf{MC}}=0.65$.  In Section \ref{section5}, ordering these simulations according to the least anisotropic to the most anisotropic simulation produced: HC, NST, MC.  We thus conjecture from the results in FIG.~\ref{extremepdf}(a) and (b) that faster dispersion linked to anisotropy will lead to a higher value of the scale parameter.

\section{Discussion \label{section7}}

We have shown that the convex hull can be used to characterize
many-particle dispersion in turbulent flows, and can reproduce
scalings similar to particle-pair, and other
multi-particle Lagrangian statistics.  The convex hull allows us to
extract dispersion behaviors that produce clear scalings from groups
of tracer particles that are significantly larger than have been
typically examined by multi-particle statistics.  We have examined
particle dispersion using convex hulls across three types of
physically distinct turbulence simulations, including Navier-Stokes
turbulence, Boussinesq convection, and MHD Boussinesq convection.  In
each of the simulations that we consider, we have shown that the
convex hull describes well the dynamics of the entire group of
particles. In addition, these tests yield further information about the turbulent
velocity field by quantifying the dynamical differences between interior particles and convex hull vertices.   
Dispersion curves produced using the maximal ray of the
convex hull, the surface area of the convex hull, and the volume of
the convex hull produce ballistic and diffusive scalings, which can be
compared with particle-pair dispersion curves. 
Although the convex
hull has been used to calculate volumes occupied by particles in some
specialized contexts \citep{dietzel2013numerical, lakes}, this is the
first time that the convex hull of the positions of Lagrangian tracer
particles has been used as a fundamental diagnostic to obtain
Lagrangian statistics of multi-particle dispersion in homogeneous
turbulent flows.

In addition, we have explored the convex
  hull's fundamental link to extreme value statistics.  We have
  discussed that the convex hull provides new information about
  extremes of dispersion that standard multi-particle statistics
  cannot.   Convex hulls
  calculated from large numbers of particles provide an ideal
  application for extreme value theory, an asymptotic theory for large
  samples.  Predictions based on extreme value theory are of practical
  use for studies of contaminants or of energetic particles, where
  questions about maximal dispersion are critical.
 Experimentally it may be simpler to track the convex hull of a large number of particles than to track all the particles in the group individually.
    We show that the
  distribution of the square length of the maximal ray of the convex
  hull is the Gumbel case of generalized extreme value distributions.
  In addition we show that for a system that is anisotropic because of
  MHD convection, the maximal ray in each physically distinct
  direction is described well by the Gumbel distribution.  Because the
  Gumbel distribution has been successful in predicting avalanches,
  extreme rainfall, and extreme winds, this nontrivial new observation
  will provide new physical intuition for modeling anomalous
  dispersion.

In a second application of the convex hull analysis, we exploit the
relationship between convex hull surface area and volume to examine
the degree of anisotropy present in a turbulent convective flow.  Our
results reveal the extent of spatial variation of anisotropy. Moreover, 
this quantity also exhibits a probability distribution that has the same
universal shape for all three considered physical systems.  Convex
hull analysis can easily isolate dispersive characteristics in any
local region of interest, for example a region where a magnetic
structure, or strong convective plume is present.  Used in this way,
they provide a versatile supplement to standard Lagrangian
multi-particle statistics in complex turbulent flows.  Because
 of these advantages, further investigation of the convex hull to analyze
 many-particle turbulent dispersion is justified.

\begin{acknowledgements}
{\small 
We thank Luca Biferale for his helpful comments on this work.  The research leading to these results has received funding from the European Research
Council under the European Union's Seventh Framework (FP7/2007-2013)/ERC grant
agreement no. 320478.  This work has also been supported by the Max-Planck
Society in the framework of the Inter-institutional
Research Initiative ``Turbulent Transport and Ion
Heating, Reconnection and Electron Acceleration in
Solar and Fusion Plasmas'' of the MPI for Solar System
Research, Katlenburg-Lindau, and the Institute
for Plasma Physics, Garching (project MIFIF-A-AERO8047).  Original simulations were performed on the VIP, VIZ, and HYDRA computer systems at the Rechenzentrum Garching of the Max Planck Society.  Additional calculations were performed on the Konrad and Gottfried computer systems of the Norddeutsche Verbund zur F\"orderung des Hoch- und H\"ochstleistungsrechnens (HLRN).
NWW acknowledges travel support from KLIMAFORSK project number 229754 and the London Mathematical Laboratory,  and Office of Naval Research NICOP grant NICOP - N62909-15-1-N143 at Warwick and Potsdam.
}
\end{acknowledgements}

\appendix
\section{Dependence of convex hull statistics on initial size and density \label{appendixsizeden}}

Because the initial separation between a pair of tracer particles affects two-particle dispersion, the initial length scale $\ell_\mathrm{hull}$ of a group of tracer particles may also play a role in a convex hull analysis of many particle dispersion.
The initial density of tracer particles clearly also is significant for dispersion, because this directly determines the resolution of the convex hull surface area and volume.  For simulations NST, HC, and MC,  the initial density of tracer particles is between $0.001$ -- $0.02~\mathsf{particles}/ \eta_{\mathsf{kol}}^3$.  This density severely limits the $\ell_\mathrm{hull}$ that can be explored in these simulations.  The initial size of the particle groups is chosen to be, $20 \eta_{\mathsf{kol}} \leq \ell_{\mathrm{hull}} \leq 30 \eta_{\mathsf{kol}}$, and
groups with substantially smaller $\ell_\mathrm{hull}$ clearly do not contain enough particles for the convex hull to be adequately resolved.
For a point of comparison, the particle groups examined by \citet{bianchi2016evolution} are significantly smaller and more dense; they have initial length scale of $\ell_{\mathrm{hull}} = \eta_{\mathsf{kol}}$ which contains 2000 tracer particles.  

In order to systematically test the convex hull analysis of dispersion for a range of initial sizes, we perform a test simulation of forced
homogeneous isotropic Navier-Stokes turbulence, similar to simulation NST, in which tracer particles are initialized in groups with given $\ell_\mathrm{hull}$, 
and at two different fixed particle densities.  The first density,  $\rho_{\mathsf{low}} =  0.005 ~\mathsf{particles}/ \eta_{\mathsf{kol}}^3$, is selected to be similar to the tracer particle density in simulations NST, HC, and MC, in order to examine $\ell_\mathrm{hull}$ both larger and smaller those examined in these simulations.  At this density we examine groups of tracer particles
with 8 initial sizes between $14 \eta_{\mathsf{kol}} \leq \ell_{\mathrm{hull}} \leq 64 \eta_{\mathsf{kol}}$.  The second density $\rho_{\mathsf{hi}} = 11.5 ~\mathsf{particles}/ \eta_{\mathsf{kol}}^3 $
is significantly higher, so that we can examine groups of particles with four smaller initial sizes between $4 \eta_{\mathsf{kol}} \leq \ell_{\mathrm{hull}} \leq 14 \eta_{\mathsf{kol}}$.

Regardless of $\ell_\mathrm{hull}$ and initial particle density, the trends evident for the convex hull validity diagnostic in FIGs.~\ref{newclumpgraph1} and \ref{newclumpgraph2} are recovered. The time at which the normalized average difference between centers $\delta c = \langle|\vec{c}_{\mathsf{vtx}}-\vec{c}_{\mathsf{int}}|/ d \rangle$ reaches a peak appears to be approximately independent of both the $\ell_\mathrm{hull}$ and density.  For larger $\ell_\mathrm{hull}$ the growth of $\delta c$ begins earlier, although the time at which $\delta c$ begins to grow is not directly relatable to $\tau_0$.  The decrease to a plateau, evident in FIG.~\ref{newclumpgraph2}, is larger for groups of particles with larger $\ell_\mathrm{hull}$, at fixed particle density for both densities tested.

\begin{figure}[H]
 \resizebox{3.45in}{!}{\includegraphics[angle=90]{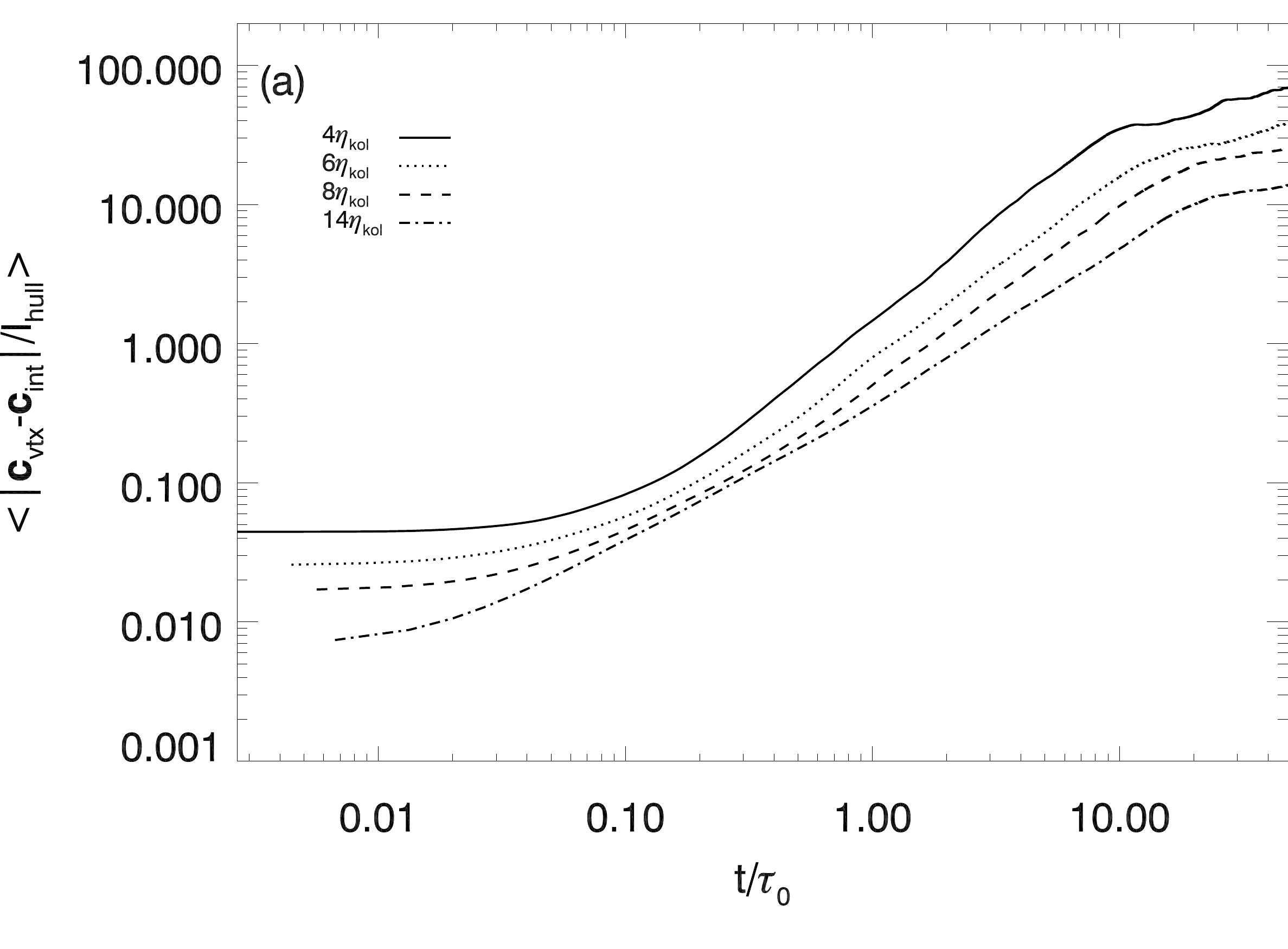}}
 \resizebox{3.45in}{!}{\includegraphics[angle=90]{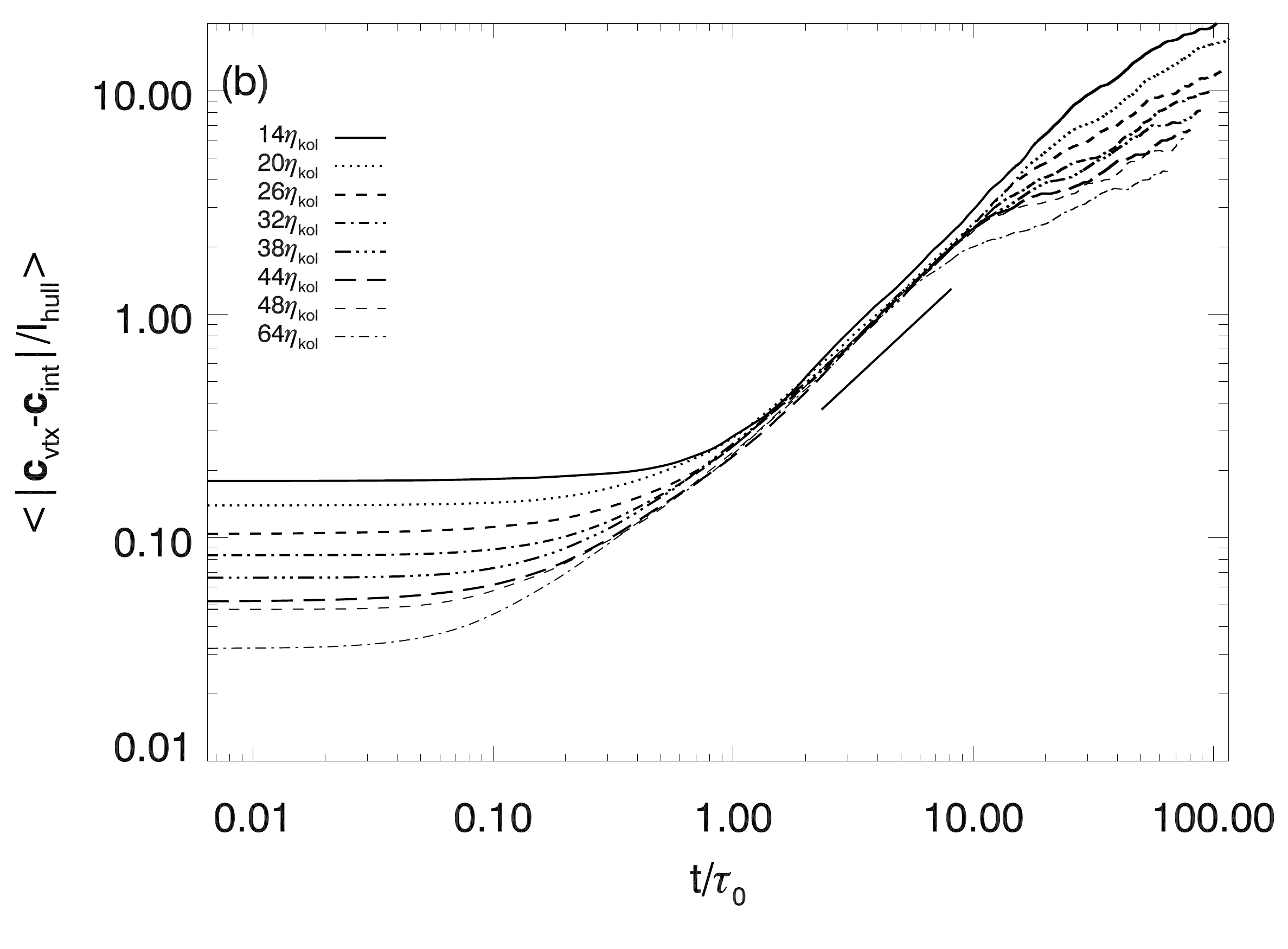}}
\caption{Evolution of the difference between the geometric center of the convex hull and the center of mass of the group of tracer particles, normalized by the initial length scale of the convex hull $\langle|\vec{c}_{\mathsf{vtx}}-\vec{c}_{\mathsf{int}}|/ \ell_{\mathrm{hull}} \rangle$ for (a) groups of particles with four initial sizes and particle density $\rho_{\mathsf{hi}}$, and (b) groups of particles with eight initial sizes and particle density $\rho_{\mathsf{low}}$.  The initial size $\ell_\mathrm{hull}$ is labeled in units of $\eta_{\mathsf{kol}}$.  A thin solid line with slope 1 is indicated during time scales associated with the inertial range.
\label{centerclumpgraph} } 
\end{figure} 
Aside from these diagnostics, we consider the growth of the difference between the geometric center of the convex hull and the center of mass of the group of tracer particles, normalized by the \emph{initial} length scale of the convex hull $\langle|\vec{c}_{\mathsf{vtx}}-\vec{c}_{\mathsf{int}}|/ \ell_{\mathrm{hull}} \rangle$.  This is not useful for examining whether the convex hull describes the group of particles well, because unlike the maximal extent $d$, $\ell_{\mathrm{hull}}$ describes the initial state of the particle group and does not change in time; however the difference in these centers provides a new quantity linked to the dispersion of many tracer particles.  The evolution of this quantity is shown in FIG.~\ref{centerclumpgraph}.   In the figure, the magnitude of the difference in the centers is clearly linked to $\ell_\mathrm{hull}$, with larger $\ell_\mathrm{hull}$ leading to smaller values of $\langle|\vec{c}_{\mathsf{vtx}}-\vec{c}_{\mathsf{int}}|/ \ell_{\mathrm{hull}} \rangle$ throughout dispersion.
However the shape of the evolution curves for the difference in the centers appears to be approximately independent of $\ell_\mathrm{hull}$.  During the inertial range of time scales, the difference in centers grows linearly with time.  Explaining this interesting scaling result will be the subject of future work.

\begin{figure}[H]
 \resizebox{3.45in}{!}{\includegraphics[angle=90]{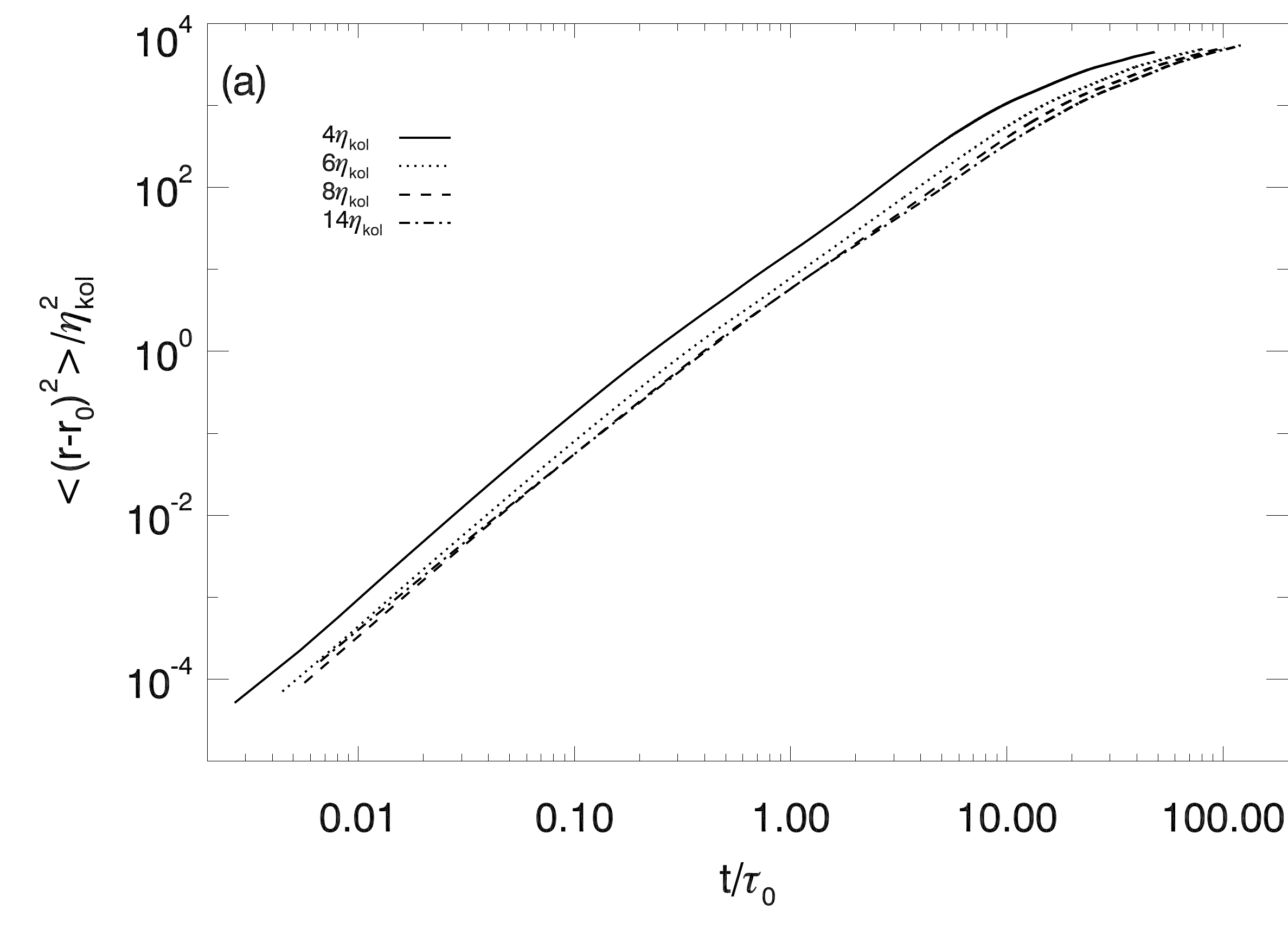}}
 \resizebox{3.45in}{!}{\includegraphics[angle=90]{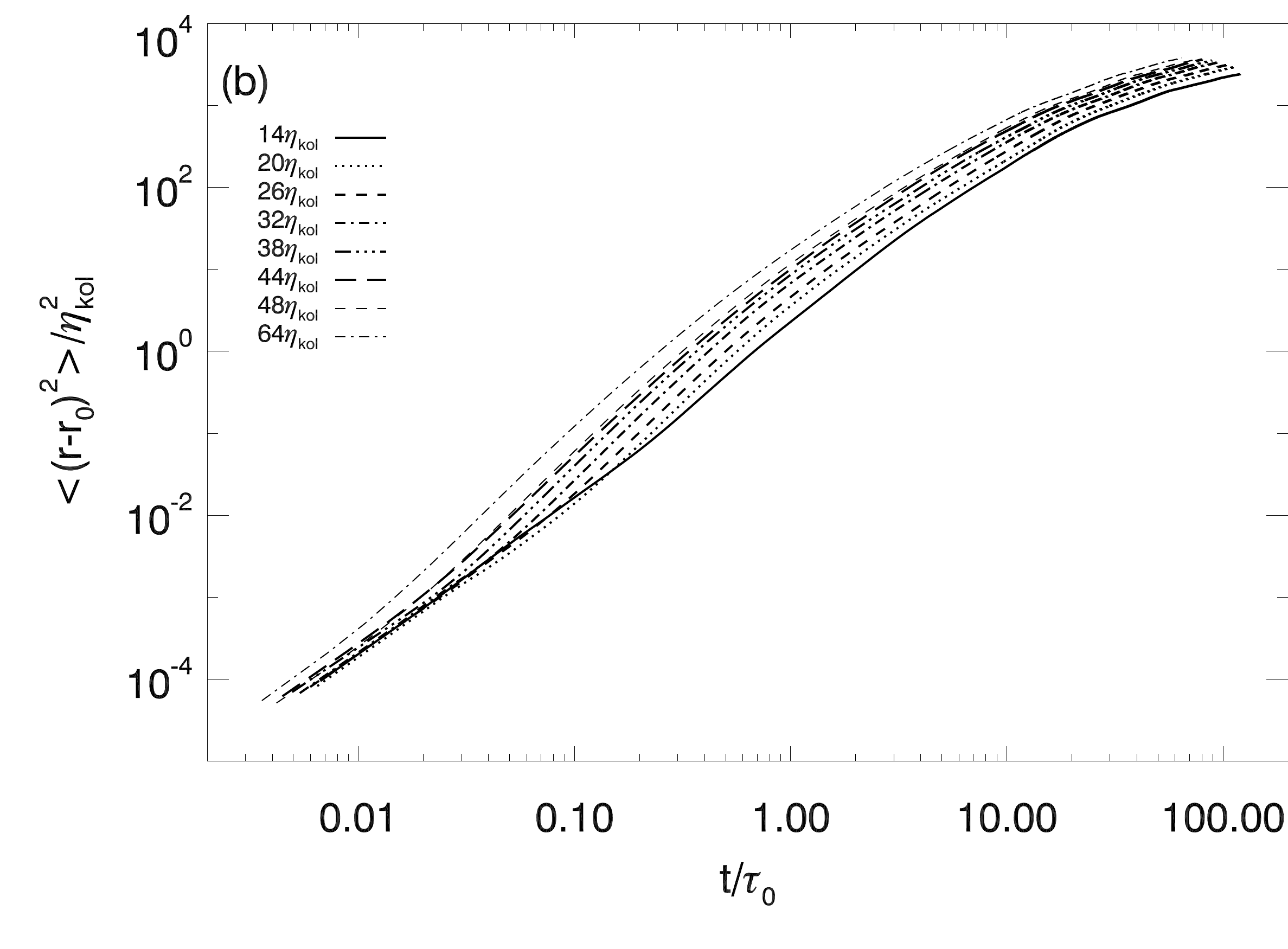}}
\caption{Evolution of the mean-square maximal ray $r$ for groups of particles with (a) four initial sizes and particle density $\rho_{\mathsf{hi}}$, and (b) eight initial sizes and particle density $\rho_{\mathsf{low}}$.  The initial size $\ell_\mathrm{hull}$ is labeled in units of $\eta_{\mathsf{kol}}$.   Thin solid lines with slope 2 and slope 1 indicate the ballistic and diffusive regimes respectively.
\label{rayclumpgraph} } 
\end{figure}
The evolution of the maximal ray $r$ introduced in Section~\ref{secmaxray} universally exhibits a clear diffusive regime for all $\ell_\mathrm{hull}$ and densities tested.   This is illustrated by FIG.~\ref{rayclumpgraph}.   We do not expect perfect ballistic scaling of the maximal ray, because unlike a particle pair, the particles that determine the maximal ray of a convex hull can be exchanged.  Despite this, we do observe a scaling reminiscent of ballistic behavior for all $\ell_\mathrm{hull}$ and densities tested; this likely indicates that vertex exchange is not a dominant effect during this early regime of dispersion.  The slope
of the dispersion curves during transitional regimes between ballistic and diffusive appears to be dependent on the initial length scale $\ell_\mathrm{hull}$.  This is unsurprising because it is a well-known result for particle-pairs.

\bibliographystyle{apsrev4-1}

\bibliography{postdoc}

\end{document}